\documentclass{hnp06}

\usepackage{bm}
\usepackage[fleqn]{amsmath}
\usepackage{amssymb}

\newcommand{\Slash}[1]{\ooalign{\hfil/\hfil\crcr$#1$}}
\newcommand{\re}{\text{Re}}

\newcommand{\bra}[1]{\langle \, #1 \, |}
\newcommand{\ket}[1]{| \, #1 \, \rangle}
\newcommand{\piK}{\pi^- p\to K^-\Theta^+}
\newcommand{\Kpi}{K^+ p\to \pi^+\Theta^+}
\newcommand{\tenbar}{\overline{\bm{10}}}

\begin{document}
\setcounter{page}{1}

\title{Phenomenological study for the $\Theta^+$ and two-meson coupling}

\author{Tetsuo Hyodo$^*$ and Atsushi Hosaka}

\address{Research Center for Nuclear Physics (RCNP)
Ibaraki 567--0047, Japan\\ 
E-mail: hyodo@yukawa.kyoto-u.ac.jp}

\address{$^*$ Present address: Yukawa Institute for Theoretical Physics,
Kyoto University, Kyoto 606--8502, Japan}

\maketitle

\abstracts{
We examine several assignments of spin and parity for the pentaquark 
$\Theta^+$ state ($J^P=1/2^{\pm}, 3/2^{\pm}$) in connection with 
phenomenology of known baryon resonances, using a general framework based on
the flavor symmetry. Assuming that the $\Theta^+$ belongs to an antidecuplet
representation which mixes with an octet, we calculate the mass spectra of 
the flavor partners of the $\Theta^+$ based on the SU(3) symmetry. The decay
widths of the $\Theta^+$ and nucleon partners are analyzed for the 
consistency check of the mixing angle obtained from the masses. It is found 
that a suitable choice of the mixing angle successfully reproduces the 
observed masses of exotics, when their spin and parity are assigned to be 
$J^P=3/2^-$, together with other nonexotic resonances of $J^P=3/2^-$. The 
decay widths of  $\Theta\to KN$, $N(1520)\to \pi N$, and $N(1700)\to \pi N$ 
are also reproduced simultaneously. 
We then evaluate two-meson couplings of $\Theta^+$, using experimental 
information of nucleon partners decaying into $\pi\pi N$ channels, in which 
the two pions are in scalar- and vector-type correlations. We examine two 
assignments of spin and parity $J^P=1/2^+$ and $3/2^-$, for which the 
experimental spectra of known resonances with exotic baryons are properly 
reproduced by an octet-antidecuplet representation mixing scheme. Using the 
obtained coupling constants, total cross sections of the reactions $\pi^- p 
\to K^- \Theta^+$ and $K^+ p\to \pi^+ \Theta^+$ are calculated. Substantial
interference of two terms may occur in the reaction processes for the 
$J^P=1/2^+$ case, whereas the interference effect is rather small for the 
$3/2^-$ case. 
}

% % % % % % % % % % % % % % % % % % % % % 

%%%%%%%%%%%%%%%%%%%%%%%%%%%%%%%%%%%%%%%%%%%%%%%%%%%%%%%%%%%%%%%%%%%%%%%%
\section{Introduction}\label{sec:Intro}
%%%%%%%%%%%%%%%%%%%%%%%%%%%%%%%%%%%%%%%%%%%%%%%%%%%%%%%%%%%%%%%%%%%%%%%%

In this article, we would like to discuss two topics; one is the 
representation mixing scheme to study the properties of the exotic state
with the flavor SU(3) symmetry\cite{Hyodo:2005wa}, and the other is the 
two-meson coupling of the $\Theta^+$ based on the mixing 
scheme\cite{Hyodo:2005bw}. The former is reported in 
section~\ref{sec:mixing} and the latter is discussed in 
section~\ref{sec:reaction}.

% % % % % % % % % % % % % % % % % 
\vspace{0.5cm}

In order to study the exotic particles, it is important to consider 
simultaneously other members with nonexotic flavors in the same SU(3) 
multiplet which the exotic particles belong to. The identification of the 
flavor multiplet provides the foundation of the model calculation, for 
instance, in order to construct effective interaction Lagrangians, or to 
construct wave functions in the constituent quark models. In addition, when 
we successfully determine the flavor multiplet, the spin and parity $J^P$ of
exotic states can be specified by the identified non-exotic partners whose 
$J^P$ is known. Since the SU(3) flavor symmetry with its breaking governs 
the relations among hadron masses and interactions\cite{deSwart:1963gc}, we 
naively expect that the exotic states also follows the symmetry relations. 
In other words, the existence of exotic particles would require the flavor 
partners, if the flavor SU(3) symmetry plays the same role as in the 
ordinary three-quark baryons.

In this study, we assume that $\Theta(1540)$\cite{Nakano:2003qx} and 
$\Xi_{3/2}(1860)$\cite{Alt:2003vb} do exist at these energies, despite the 
controversial situation of experiments. Although we use these specific mass 
values, the symmetry relations we derive here are rather general, and can 
be applied to other exotic states once they are assumed to belong to the 
same SU(3) representations. Therefore, when any other exotic particles (with
the quantum number of $\Theta^+$ or $\Xi_{3/2}$) are found experimentally in
future, we can immediately apply the present formulae to these states.

Concerning the representation that the $\Theta^+$ belongs to, there are 
several conjectures in model calculations. In the chiral soliton 
models\cite{Diakonov:1997mm}, the $\Theta^+$ and $\Xi_{3/2}$ belong to the 
antidecuplet ($\overline{\bm{10}}$) representation with spin and parity 
$J^P=1/2^+$. An interesting proposal was made by Jaffe and 
Wilczek\cite{Jaffe:2003sg} in a quark model with diquark correlation. The 
model is based on the assumption of the strong diquark correlation in 
hadrons and the representation mixing of an octet ($\bm{8}$) with an 
antidecuplet ($\overline{\bm{10}}$). The attractive diquark correlation in 
the scalar-isoscalar channel leads to $J^P=1/2^+$ for the $\Theta^+$. With 
the ideal mixing of $\bm{8}$ and $\overline{\bm{10}}$, in which states are 
classified by the number of strange and anti-strange quarks, $N(1710)$ and 
$N(1440)$ resonances are well fit into members of the multiplet together 
with the $\Theta^+$. However, it was pointed out that mixing angles close 
to the ideal one encountered a problem in the decay pattern of $N(1710)\to 
\pi N$ and $N(1440)\to \pi N$. Rather, these decays implied a small mixing 
angle\cite{Cohen:2004gu,Pakvasa:2004pg,Praszalowicz:2004xh}. This is 
intuitively understood by the discrepancy between the broad decay width of 
$N(1440)\to \pi N$ and the narrow decay widths of $N(1710)\to \pi N$ and 
$\Theta\to K N$\cite{Glozman:2003vb,Cohen:2004gu}.

At this stage, it is worth noting that the flavor SU(3) symmetry itself 
does not constrain the spin and parity. Therefore, employing the 
$\bm{8}$-$\overline{\bm{10}}$ mixing scenario which is the minimal scheme to
include the $\Theta^+$ and $\Xi^{--}$, here we examine the possibilities to 
assign other quantum numbers, such as $1/2^-$, $3/2^+$, $3/2^-$, and search 
for proper nucleon partners of the exotic states among the known resonances.
Although the mass formulae were already given previously, they are applied 
mainly to the $J^P=1/2^+$ case in accordance with the Jaffe-Wilczek 
model\cite{Jaffe:2003sg}, and sometimes to the $J^P=1/2^-$. The spin 
$3/2$ states are rarely examined. This is natural because the lower spin 
states are expected to be lighter. However, once again, the flavor symmetry
is nothing to do with the spin and parity by itself, therefore we 
investigate the $J^P=3/2^{\pm}$ states as well. Indeed, we find a natural 
solution consistent with both the masses and widths in the $3/2^-$ case. 

The present study is based on the flavor SU(3) symmetry, experimental mass 
spectra and decay widths of the $\Theta^+$, the $\Xi^{--}$ and known baryon 
resonances. Hence, the analysis is rather phenomenological, but does not 
rely upon any specific models. For instance, we do not have to specify the 
quark contents of the baryons. Although the exotic states require minimally 
five quarks, nonexotic partners do not have to. Instead, we expect that the 
resulting properties such as masses and decay rates reflect information from
which we hope to learn internal structure of the baryons. 

% % % % % % % % % % % % % % % % % 
\vspace{0.5cm}

Now, a particularly interesting property that is expected to be 
characteristic for exotic baryons is their strong coupling to two-meson 
states in transitions to an ordinary baryon, as studied in Refs.~12,13. 
Studying two-meson couplings of the exotic baryon $\Theta^+$ is important 
for the following reasons. 

First, a heptaquark model has been proposed in the early stage of the study 
of the $\Theta^+$ to explain a light mass and a narrow decay 
width\cite{Bicudo:2003rw,Kishimoto:2003xy,Llanes-Estrada:2003us,Bicudo:2004cm,Huang:2004ti}. 
Although a quantitative study---in particular with a model of hadrons where 
$\Theta^+$ is regarded as a bound state of $\pi K N$ system---does not make 
self-bound system with the present knowledge of hadron interactions, a 
two-meson contribution to the self-energy of $\Theta^+$ has been shown to be
consistent with the expected pattern of the masses of the antidecuplet
members\cite{Hosaka:2004mv}.

Second, the importance of two-meson coupling has been implied from an
empirical observation of the generalized OZI rule\cite{Roy:2003hk}. The
dominance of connected quark lines favors creation of a $q \bar q$ pair in 
the transition of $\Theta^+(qqqq\bar q) \to N(qqq)$, which is naturally 
associated with couplings to two mesons. On the other hand, couplings to a 
single meson, which are called as ``fall apart'' decay, should be 
suppressed, since the final states are not connected by the quark line.

Finally, two-meson couplings play important roles in reaction studies.
Without two-meson couplings, all the amplitudes for $\Theta^+$ production 
are proportional to the $\Theta^+KN$ coupling, which is fixed by the very 
small decay width of the $\Theta^+$. However, two-meson couplings are 
determined from other sources as we will see in the following, independently
of the $\Theta^+KN$ coupling. Therefore, even with the extremely narrow 
width of $\Theta^+$, a sizable cross section can be obtained with two-meson 
couplings.
 
In Ref.~13, an analysis of the two-meson coupling was performed in the study
of the self-energy of the $\Theta^+$, assuming that $J^P=1/2^+$ with 
$N(1710)\equiv N^*$ in the same antidecuplet ($\bm{\overline{10}}$). Since 
the $\Theta^+$ cannot decay into $K\pi N$ channel, the coupling constants 
were determined from the $N^*$ decay into the $\pi\pi N$ channel and flavor 
SU(3) symmetry. Two types of Lagrangians were found to be important for the 
self-energy of the baryon antidecuplet. It was also shown that the two-meson
contribution was indeed dominant over a single-meson contribution. However, 
the assumption of pure $\bm{\overline{10}}$ for $\Theta$ and $N^*$ may not 
be the case in reality.

Now we can improve this point based on the study of the representation 
mixing, where we find reasonable fits for $J^P=1/2^+$ and $3/2^-$. 
Therefore, here we would like to calculate the two-meson couplings including
the representation mixing. First we determine the coupling constants of 
$N^*\to \pi\pi N$ from the experimental widths and separate the 
$\bm{\overline{10}}$ component from the $\bm{8}$ component. Then, by using 
SU(3) symmetry, the coupling constants of $\Theta  K\pi N$ are determined 
for $J^P=1/2^+$ and $3/2^-$, including representation mixing of $\bm{8}$ and
$\bm{\overline{10}}$. We focus on the decay channels in which the two pions 
are correlated in scalar-isoscalar and vector-isovector channels, which are 
the main decay modes of the resonances and play a dominant role in the 
$\Theta^+$ self-energy\cite{Hosaka:2004mv}.

As an application of the effective interaction of $\Theta K\pi N$, we 
perform the analysis of the $\piK$ and $\Kpi$ reactions. These reactions 
were studied using effective Lagrangian 
approaches\cite{Hyodo:2003th,Liu:2003rh,Oh:2003gj,Oh:2003kw}. Experiments 
for $\piK$ have been performed at KEK\cite{Miwa:2006if}, and a 
high-resolution experiment for the $\Kpi$ reaction is ongoing. We can 
compare the results with these experiments, which may help to determine the 
$J^P$ of the $\Theta^+$.

%%%%%%%%%%%%%%%%%%%%%%%%%%%%%%%%%%%%%%%%%%%%%%%%%%%%%%%%%%%%%%%%%%%%%%%%
\section{Flavor multiplet of the $\Theta^+$}\label{sec:mixing}
%%%%%%%%%%%%%%%%%%%%%%%%%%%%%%%%%%%%%%%%%%%%%%%%%%%%%%%%%%%%%%%%%%%%%%%%

%%%%%%%%%%%%%%%%%%%%%%%%%%%%%%%%%%%%
\subsection{Analysis with pure antidecuplet}\label{subsec:bar10}

First we briefly discuss the case where the $\Theta^+$ belongs to pure 
$\overline{\bm{10}}$ without mixing with other representations. In this 
case, the masses of particles belonging to the $\overline{\bm{10}}$ can be 
determined by the Gell-Mann--Okubo (GMO) mass formula
\begin{equation}
    M(\overline{\bm{10}};Y)
    \equiv \bra{\overline{\bm{10}};Y}\mathcal{H}
    \ket{\overline{\bm{10}};Y}
    =  M_{\overline{\bm{10}}} - aY  ,
    \label{eq:bar10mass_1}
\end{equation}
where $Y$ is the hypercharge of the state, and $\mathcal{H}$ denotes the 
mass matrix. Note that at this point the spin and parity $J^P$ are not yet
specified. The quantum numbers will be assigned as explained below. 

In Eq.~\eqref{eq:bar10mass_1}, there are two parameters, 
$M_{\overline{\bm{10}}}$ and $a$, which are not determined by the flavor 
SU(3) symmetry. However, we can estimate the values of these parameters by 
considering their physical meaning in some models. For instance, in a 
constituent quark model, $\overline{\bm{10}}$ can be minimally expressed as 
four quarks and one antiquark. Therefore, $M_{\overline{\bm{10}}}$ should be
at least larger than the masses of three-quark baryons. In this picture, the
mass difference of $\Xi(ssqq\overline{q})$ and $\Theta(qqqq\overline{s})$, 
namely $3a$, is provided by the difference between the constituent masses of
the light ($ud$) and strange quarks, which is about 100-250 
MeV\cite{Hosaka:2004mv}. On the other hand, in the chiral quark soliton 
model, $3a$ is related to the pion nucleon sigma 
term\cite{Schweitzer:2003fg}. In this picture $3a$ can take values in the 
range of 300-400 MeV, due to the experimental uncertainty of the pion 
nucleon sigma term $\Sigma_{\pi N}=$64-79 
MeV\cite{Diakonov:2003jj,Ellis:2004uz}. Note that in the chiral soliton 
model, spin and parity are assigned as $J^P=1/2^+$ for the antidecuplet.

Taking the above estimation into consideration, we test several parameter 
sets fixed by the experimentally known masses of particles. The results are 
summarized in Table~\ref{tbl:bar10result}. We use the mass of the $\Theta^+$
$M_{\Theta}=1540$ MeV and pick up a possible nucleon resonance. Assuming 
that the $\Theta^+$ and the nucleon resonance are in the same SU(3) 
multiplet, the $J^P$ of the multiplet is determined by that of the nucleon 
partner. In the 
three cases of $J^P=1/2^+, 3/2^{\pm}$, the exotic $\Xi$ resonance is 
predicted to be higher than 2 GeV, and the inclusion of $\Xi(1860)$ in the 
same multiplet seems to be difficult. Furthermore, the $\Sigma$ states 
around 1.8-1.9 GeV are not well assigned (either two-star for $J^P=1/2^+$, 
or not seen for $J^P=3/2^{\pm}$). Therefore, fitting the masses in the pure 
antidecuplet scheme seems to favor $J^P=1/2^-$.

%--table----------------------------
\begin{table}[btp]
    \centering
    \caption{Summary of subsection~\ref{subsec:bar10}. Masses and $\Theta^+$
    decay widths are shown for several assignments of quantum numbers. For 
    $1/2^-$ the masses of $\Theta$ and $\Xi_{3/2}$ are the input parameters,
    while for $1/2^+,3/2^{\pm}$, the masses of $\Theta$ and $N$ are the 
    input parameters. Values in parenthesis are the predictions, and we show
    the candidates to be assigned for the states. All values are listed in 
    units of MeV.}
    \vspace{0.2cm}
    \begin{tabular}{|c|cccc|l|}
	\hline
	$J^P$ & $M_{\Theta}$ & $M_{N}$ & $M_{\Sigma}$ & $M_{\Xi}$
	& $\Gamma_{\Theta}$ \\
	\hline
	& 1540 & [1647] & [1753] & 1860   &    \\
	$1/2^-$ & & $N(1650)$ & $\Sigma(1750)$ &  & 
	$156.1 \ {}^{+90.8}_{-73.3}$  \\
	&  1540  & 1710  & [1880] &[2050] & \\
	$1/2^+$ & &  & $\Sigma(1880)$ & $\Xi(2030)$ &
	  $\phantom{00}7.2 \ {}^{+15.3}_{-4.6}$  \\
	& 1540  & 1720  & [1900] & [2080] &   \\
	$3/2^+$  &  &  & - & - & $\phantom{0}10.6 \ {}^{+7.0}_{-5.0}$  \\
	& 1540 & 1700  & [1860] & [2020] &   \\
	$3/2^-$ & &  & - & $\Xi(2030)$ 
	 & $\phantom{00}1.3 \ {}^{+1.2}_{-0.9}$  \\
	 \hline
    \end{tabular}
    \label{tbl:bar10result}
\end{table}
%--table----------------------------

Next we study the decay widths of the $N^*$ resonances with the above 
assignments. For the decay of a resonance $R$, we define the dimensionless 
coupling constant $g_R$ by
\begin{equation}
    \Gamma_R\equiv 
    g_R^{2}F_I \frac{p^{2l+1}}{M_R^{2l}} ,
    \label{eq:coupling}
\end{equation}
where $p$ is the relative three momentum of the decaying particles in the 
resonance rest frame, $l$ is the angular momentum of the decaying particles,
$\Gamma_R$ and $M_R$ are the decay width and the mass of the resonance $R$. 
$F_I$ is the isospin factor (2 for $\Theta\to KN$ and 3 for $N^*\to \pi N$).
Assuming flavor SU(3) symmetry, a relation between the coupling constants of
$\Theta \to K N$ and $N^*\to \pi N$ is given by:
\begin{equation}
    g_{\Theta KN}=\sqrt{6}g_{N^*\pi N}  .
    \label{eq:relation}
\end{equation}
With these formulae~\eqref{eq:coupling} and \eqref{eq:relation}, we 
calculate the decay width of the $\Theta^+$ from that of $N^*\to \pi N$ of 
the nucleon partner. Results are also shown in Table~\ref{tbl:bar10result}. 
We quote the errors coming from experimental uncertainties in the total 
decay widths and branching ratios, taken from the Particle Data Group 
(PDG)\cite{Eidelman:2004wy}. It is easily seen that as the partial wave of 
the two-body final states becomes higher, the decay width of the resonance 
becomes narrower, due to the effect of the centrifugal barrier. Considering 
the experimental width of the $\Theta^+$, the results of $J^P=3/2^-$, 
$3/2^+$, $1/2^+$ are acceptable, but the result of the $J^P=1/2^-$ case, 
which is of the order of hundred MeV, is unrealistic.

In summary, it seems difficult to assign the $\Theta^+$ in the pure 
antidecuplet $\overline{\bm{10}}$ together with known resonances of 
$J^P=1/2^{\pm},3/2^{\pm}$, by analyzing the mass and width simultaneously.

%%%%%%%%%%%%%%%%%%%%%%%%%%%%%%%%%%%%
\subsection{Analysis with octet-antidecuplet mixing}\label{sec:8bar10}

In this section we consider the representation mixing between 
$\overline{\bm{10}}$ and $\bm{8}$. Here we work under the assumption of 
minimal $\bm{8}$-$\overline{\bm{10}}$ mixing. The nucleon and $\Sigma$ 
states in the $\bm{8}$ will mix with the states in 
the $\overline{\bm{10}}$ of the same quantum numbers. Denoting the mixing 
angles of the $N$ and the $\Sigma$ as $\theta_N$ and $\theta_{\Sigma}$, the 
physical states are represented as
\begin{align}
    \ket{N_1} =& \ket{\bm{8},N} \cos\theta_N
    - \ket{\overline{\bm{10}},N} \sin\theta_N  ,\nonumber \\
    \ket{N_2} =& \ket{\overline{\bm{10}},N} \cos\theta_N
    + \ket{\bm{8},N} \sin\theta_N  ,
    \label{eq:Nmixing} \\
    \ket{\Sigma_1} =& \ket{\bm{8},\Sigma} \cos\theta_\Sigma
    - \ket{\overline{\bm{10}},\Sigma} \sin\theta_\Sigma  , \nonumber  \\
    \ket{\Sigma_2} =& \ket{\overline{\bm{10}},\Sigma} \cos\theta_\Sigma
    + \ket{\bm{8},\Sigma} \sin\theta_\Sigma  .
    \label{eq:Sigmamixing}
\end{align}
To avoid redundant duplication, the domain of the mixing angles is 
restricted in $0\leq \theta < \pi/2$, and we will find solutions for  
$N_1$ and $\Sigma_1$ lighter than $N_2$ and $\Sigma_2$, respectively. 

When we construct $\overline{\bm{10}}$ and $\bm{8}$ from five quarks, the 
eigenvalues of the strange quark (antiquark) number operator $n_s$ of 
nucleon states become generally fractional. However, in the scenario of the 
ideal mixing of Jaffe and Wilczek, the physical states are given as
\begin{align}
    \ket{N_1}
    &= \sqrt{\frac{2}{3}}\ket{\bm{8},N}
    -\sqrt{\frac{1}{3}}\ket{\overline{\bm{10}},N}  ,
    \nonumber
    \\
    \ket{N_2}
    &= \sqrt{\frac{2}{3}}\ket{\overline{\bm{10}},N}
    +\sqrt{\frac{1}{3}}\ket{\bm{8},N} ,
    \nonumber
\end{align}
such that $\bra{N_1}n_s\ket{N_1}=0$ and $\bra{N_2}n_s\ket{N_2}=2$. In this 
case, the mixing angle is $\theta_N\sim 35.2^{\circ}$. In the Jaffe-Wilczek 
model, $N(1440)$ and $N(1710)$ are assigned to $N_1$ and $N_2$, 
respectively. Notice that the separation of the $s\bar{s}$ component in the
ideal mixing is only meaningful for mixing between five-quark states, while
the number of quarks in the baryons is arbitrary in the present general 
framework.

%%%%%%%%%%%%%%%%%%%%%%%%%%%%%%%%%%%%
\subsection{Mass spectrum}\label{subsec:mass}

Let us start with the GMO mass formulae for $\overline{\bm{10}}$ and 
$\bm{8}$ :
\begin{align}
    M(\overline{\bm{10}};Y)
    &\equiv \bra{\overline{\bm{10}};Y} \mathcal{H} 
    \ket{\overline{\bm{10}};Y}
    =  M_{\overline{\bm{10}}} - aY  ,
    \label{eq:bar10mass} \\
    M(\bm{8};I,Y)
    &\equiv \bra{\bm{8};I,Y} \mathcal{H} \ket{\bm{8};I,Y}
    = M_{\bm{8}} - bY + c
    \left[ I(I+1) -\frac{1}{4}Y^2\right]  ,
    \label{eq:8mass}
\end{align}
where $Y$ and $I$ are the hypercharge and the isospin of the state. Under 
representation mixing as in Eqs.~\eqref{eq:Nmixing} and 
\eqref{eq:Sigmamixing}, the two nucleons $(N_{\bm{8}},
N_{\overline{\bm{10}}})$ and the two sigma states $(\Sigma_{\bm{8}},
\Sigma_{\overline{\bm{10}}})$ mix, and their mass matrices are given by 
$2\times 2$ matrices. The diagonal components are given by 
Eqs.~\eqref{eq:bar10mass} and \eqref{eq:8mass}, while the off-diagonal 
elements are given as
\begin{equation}
    \bra{\bm{8},N}\mathcal{H}\ket{\overline{\bm{10}},N}
    =\bra{\bm{8},\Sigma}\mathcal{H}\ket{\overline{\bm{10}},\Sigma}
   \equiv \delta  .
   \nonumber
\end{equation}
The equivalence of the two off-diagonal elements can be verified when the 
symmetry breaking term is given by $\lambda_8$ due to a finite strange 
quark mass\cite{Diakonov:2003jj}.

The physical states $\ket{N_i}$ and $\ket{\Sigma_i}$ diagonalize 
$\mathcal{H}$. Therefore, we have the relations
\begin{equation}
    \tan 2\theta_N
    = \frac{2\delta}{M_{\overline{\bm{10}}}
    -M_{\bm{8}}-a+b-\frac{1}{2}c},
    \quad
    \tan 2\theta_{\Sigma}
    = \frac{2\delta}{M_{\overline{\bm{10}}}
    -M_{\bm{8}}-2c}. 
    \nonumber
\end{equation}
Now we have the mass formulae for the states
\begin{align}
    M_{\Theta}
    =& M_{\overline{\bm{10}}}-2a  ,
    \nonumber
    \\
    M_{N_1}
    =& \left(M_{\bm{8}}-b+\frac{1}{2}c\right)\cos^2\theta_N
    +\left(M_{\overline{\bm{10}}}-a\right)\sin^2\theta_N 
    -\delta \sin 2\theta_N ,
    \nonumber
    \\
    M_{N_2}
    =& \left(M_{\bm{8}}-b+\frac{1}{2}c\right)\sin^2\theta_N
    +\left(M_{\overline{\bm{10}}}-a\right)\cos^2\theta_N
    +\delta \sin 2\theta_N   ,  
    \nonumber
    \\
    M_{\Sigma_1}
    =& \left(M_{\bm{8}}+2c\right)\cos^2\theta_{\Sigma}
    +M_{\overline{\bm{10}}}\sin^2\theta_{\Sigma}
    -\delta \sin 2\theta_{\Sigma} ,
    \nonumber \\
    M_{\Sigma_2}
    =& \left(M_{\bm{8}}+2c\right)\sin^2\theta_{\Sigma}
    +M_{\overline{\bm{10}}}\cos^2\theta_{\Sigma}
    +\delta \sin 2\theta_{\Sigma} ,
    \nonumber \\
    M_{\Lambda}
    =&M_{\bm{8}} ,
    \nonumber \\
    M_{\Xi_8}
    =&M_{\bm{8}}+b+\frac{1}{2}c ,
    \nonumber 
    \\
    M_{\Xi_{\overline{\bm{10}}}}
    =&M_{\overline{\bm{10}}}+a .
    \nonumber
\end{align}
We have altogether six parameters $M_{\bm{8}}$, $M_{\overline{\bm{10}}}$, 
$a$, $b$, $c$ and $\delta$ to describe 8 masses. Eliminating the mixing 
angles and $\delta$, we obtain a relation independent of the mixing 
angle\cite{Diakonov:2003jj}
\begin{equation}
    2(M_{N_1}+M_{N_2}+M_{\Xi_8})
    = M_{\Sigma_1}+M_{\Sigma_2}+3M_{\Lambda}+M_{\Theta} .
    \nonumber
\end{equation}

\begin{figure*}[tbp]
    \centering
    \includegraphics[width=12cm,clip]{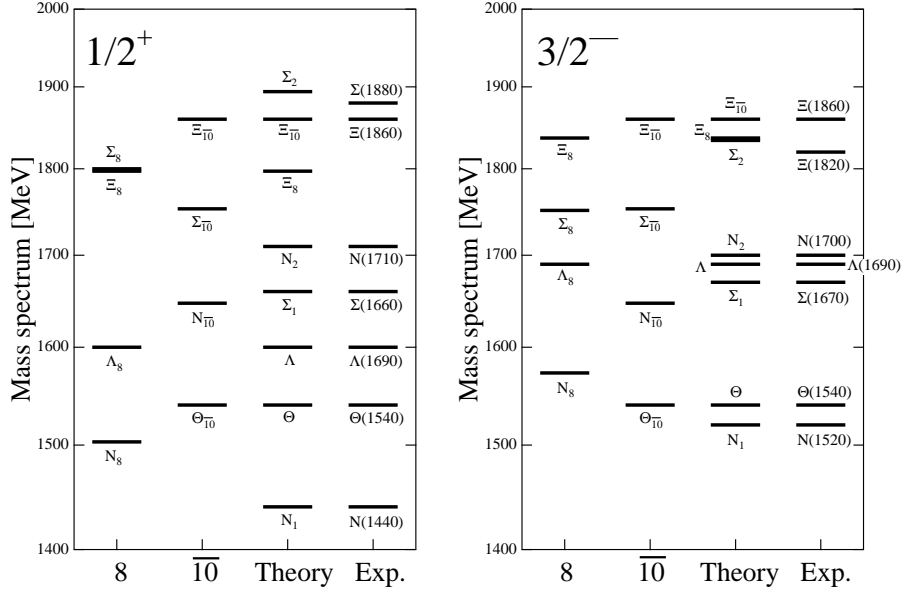}
    \caption{Results of mass spectra with representation mixing. Theoretical
    masses of the octet, antidecuplet, and the one with mixing are compared 
    with the experimental masses. In the left panel, we show the results 
    with $J^P=1/2^+$, while the results with $J^P=3/2^-$ are presented in 
    the right panel.}
    \label{fig:spectrum}
\end{figure*}

Let us first examine the case of $J^P=1/2^+$. Fixing $\Theta(1540)$,
$N_1(1440)$, $N_2(1710)$, $\Lambda(1600)$, $\Sigma_1(1660)$, 
$\Xi_{\overline{10}}(1860)$, we obtain the parameters as given in 
Table~\ref{tbl:param1}. The resulting mass spectrum together with the two 
predicted masses, $\Sigma_2=1894$ MeV and $\Xi_8=1797$ MeV, are given in 
Table~\ref{tbl:result1} and also shown in the left panel of 
Fig.~\ref{fig:spectrum}. In Fig.~\ref{fig:spectrum}, the spectra from 
experiment and those before the representation mixing are also plotted.

%--table----------------------------
\begin{table}[btp]
    \centering
    \caption{Parameters for $1/2^+$ case. All values are listed in MeV 
    except for the mixing angles.}
    \vspace{0.2cm}
    \begin{tabular}{|llllllll|}
	\hline
	 $M_{\bm{8}}$ & $M_{\overline{\bm{10}}}$ & $a$
	 & $b$ & $c$ & $\delta$
	 & $\theta_N$ & $\theta_{\Sigma}$  \\
	\hline
	 1600 & 1753.3 & 106.7 & 146.7 & 100.1 & 114.4
	& $29.0^{\circ}$ & $50.8^{\circ}$ \\
	\hline
    \end{tabular}
    \label{tbl:param1}
\end{table}
%--table----------------------------

%--table----------------------------
\begin{table}[btp]
    \centering
    \caption{
    Mass spectra for $1/2^+$ case. All values are listed in MeV.
    Values in parenthesis ($\Sigma_2$ and $\Xi_{\bm{8}}$)
    are predictions (those which are not 
    used in the fitting).
    }
    \vspace{0.2cm}
    \begin{tabular}{|cccccccc|}
	\hline
	$\Theta$ & $N_1$ & $N_2$ & $\Sigma_1$ & $\Sigma_2$
	& $\Lambda$ & $\Xi_{\bm{8}}$ & $\Xi_{\overline{\bm{10}}}$  \\
	\hline
	1540 & 1440 & 1710 & 1660
	& [1894]  & 1600 & [1797]  & 1860  \\
	\hline
    \end{tabular}
    \label{tbl:result1}
\end{table}
%--table----------------------------

%--figure---------------------------------
%--figure---------------------------------

As we see in Table~\ref{tbl:result1} and Fig.~\ref{fig:spectrum}, even 
without using the $\Sigma_2$ for the fitting, this state appears in the 
proper position to be assigned as $\Sigma(1880)$. Taking into account the 
experimental uncertainty in the masses, the present result can be regarded 
as the same as that in Ref.~9, where all known resonances including 
$\Sigma(1660)$ and $\Sigma(1880)$ are used to perform the $\chi^2$ fitting.
In this scheme, we need a new $\Xi$ state around 1790-1800 MeV to complete 
the multiplet, but the overall description of the mass spectrum is 
acceptable. Note that the mixing angle $\theta_N\sim 30^{\circ}$ is 
compatible with the one of the ideal mixing, if we consider the experimental
uncertainty of masses.

It is interesting to observe that in the spectrum of the octet, as shown in 
Fig.~\ref{fig:spectrum}, the $\Xi_8$ and the $\Sigma_8$ are almost 
degenerate, reflecting the large value for the parameter $c\sim$ 100 MeV,
which is responsible for the splitting of $\Lambda$ and $\Sigma$. For the 
ground state octet, Eq.~\eqref{eq:8mass} is well satisfied with $b=139.3$ 
MeV and $c=40.2$ MeV\cite{Diakonov:2003jj}. This point will be discussed 
later.

Now we examine the other cases of $J^P$. For $J^P={1/2^-}$, as we observed 
in the previous section, the pure $\overline{\bm{10}}$ assignment works well
for the mass spectrum, which implies that the mixing with $\bm{8}$ is small,
as long as we adopt $N(1650)$ and $\Sigma(1750)$ in the multiplet. Then the 
results of $1/2^-$ with the mixing do not change from the previous results 
of the pure $\overline{\bm{10}}$  assignment, which eventually lead to a 
broad width of $\Theta^+\to KN$ of order 100 MeV. Hence, it is not realistic
to assign $1/2^-$, even if we consider the representation mixing.

Next we consider the $3/2^+$ case. In this case candidate states are not 
well established. Furthermore, the states are distributed in a wide energy 
range, and sometimes it is not possible to assign these particles in the 
$\bm{8}$-$\overline{\bm{10}}$ representation scheme. Therefore, at this 
moment, it is not meaningful to study the $3/2^+$ case unless more states 
with $3/2^+$ will be observed.

Now we look at the $3/2^-$ case. In contrast to the $3/2^+$ case, there are 
several well-established resonances. Following the same procedure as before,
we first choose the following six resonances as inputs: $\Theta(1540)$,
$N_1(1700)$, $N_2(1520)$, $\Sigma(1670)$, $\Lambda(1690)$, and 
$\Xi_{3/2}(1860)$. We then obtain the parameters as given in 
Table~\ref{tbl:param2}, and predicted masses of other members are shown in 
Table~\ref{tbl:result2}. The masses of $N(1520)$ and $N(1700)$ determine the
mixing angle of nucleons $\theta_N\sim 33^{\circ}$, which is close to the 
ideal one. Interestingly, the fitting provides $M_{\Xi_8}\sim 1837$ MeV, 
which is close to the known three-star resonance $\Xi(1820)$ of $J^P=3/2^-$.
We have obtained acceptable assignments, although a new $\Sigma$ state is 
necessary to complete the multiplet in both cases. The mass spectrum is also
shown in Fig.~\ref{fig:spectrum}. We have also tried other possible 
assignments in Ref.~1.

%--table----------------------------
\begin{table}[btp]
    \centering
    \caption{Parameters for $3/2^-$ case. All values are listed in MeV 
    except for the mixing angles.}
    \vspace{0.2cm}
    \begin{tabular}{|llllllll|}
	\hline
	 $M_{\bm{8}}$ & $M_{\overline{\bm{10}}}$ & $a$
	 & $b$ & $c$ & $\delta$
	 & $\theta_N$ & $\theta_{\Sigma}$ \\
	\hline
	1690 & 1753.3 & 106.7 & 131.9 & 30.5 & 82.2
	& $33.0^{\circ}$ & $44.6^{\circ}$ \\
	\hline
    \end{tabular}
    \label{tbl:param2}
\end{table}
%--table----------------------------

%--table----------------------------
\begin{table}[btp]
    \centering
    \caption{Mass spectra for $3/2^-$ case. All values are listed in MeV.
    Values in parenthesis are predictions (those which are not used in the 
    fitting).}
    \vspace{0.2cm}
    \begin{tabular}{|cccccccc|}
	\hline
	 $\Theta$ & $N_1$ & $N_2$ & $\Sigma_1$ & $\Sigma_2$
	& $\Lambda$ & $\Xi_{\bm{8}}$ & $\Xi_{\overline{\bm{10}}}$  \\
	\hline
	1540 & 1520 & 1700 & 1670
	& [1834]  & 1690 & [1837]  & 1860  \\
	\hline
    \end{tabular}
    \label{tbl:result2}
\end{table}
%--table----------------------------

Let us briefly look at the octet and antidecuplet spectra of $1/2^+$ and 
$3/2^-$ resonances as shown in Fig.~\ref{fig:spectrum}. The antidecuplet 
spectrum is simple, since the GMO mass formula contains only one parameter 
which describes the size of the splitting. Contrarily, the octet spectrum 
contains two parameters which could reflect more information on different 
internal structure. As mentioned before, in the octet spectrum of $1/2^+$, 
the mass of $\Sigma_8$ is pushed up slightly above $\Xi_8$, significantly 
higher than $\Lambda_8$. This pattern resembles the octet spectrum which is 
obtained in the Jaffe-Wilczek model, where baryons are made with two flavor 
$\bar{\bm{3}}$ diquarks and one antiquark. In contrast, the spectrum of the 
octet of $3/2^-$ resembles the one of the ground state octet; we find the 
parameters $(b,c)=(131.9,30.5)$ MeV, which are close to 
$(b,c)=(139.3,40.2)$ MeV for the ground states. This is not far from the 
prediction of an additive quark model of three valence quarks. It would be 
interesting to investigate further the quark contents from such a different 
pattern of the mass spectrum.  

%%%%%%%%%%%%%%%%%%%%%%%%%%%%%%%%%%%%
\subsection{Decay width}

In the previous subsection, mass spectra of the $J^P=1/2^+$ and $J^P=3/2^-$ 
are reasonably well described. Here we study the consistency of the mixing 
angle obtained from mass spectra with the one obtained from nucleon decay 
widths for these two cases. Using Eq.~\eqref{eq:relation}, we define a 
universal coupling constant $g_{\overline{\bm{10}}}$ as
\begin{equation}
    g_{\Theta KN} = \sqrt{6} g_{N_{\overline{\bm{10}}} \pi N}
    \equiv g_{\overline{\bm{10}}}  .
    \nonumber
\end{equation}
The coupling constants of the $\pi N$ decay modes from the $N_{\bm{8}}$, 
$N_1$, and $N_2$ are defined as $g_{N_8}$, $g_{N_1}$, and $g_{N_2}$,
respectively. Under the representation mixing, these constants are related 
to each other by
\begin{align}
    g_{N_1}
    &=g_{N_{\bm{8}}} \cos\theta_N
    -\frac{g_{\overline{\bm{10}}}}{\sqrt{6}}\sin\theta_N
    , \nonumber \\
    g_{N_2}
    &=\frac{g_{\overline{\bm{10}}}}{\sqrt{6}}
    \cos\theta_N
    +g_{N_{\bm{8}}}\sin\theta_N ,\nonumber
\end{align}
and the constants can be translated into the decay widths through 
Eq.~\eqref{eq:coupling}. Notice, however, that we cannot fix the relative 
phase between $g_{N_{\bm{8}}}$ and $g_{\overline{\bm{10}}}$. Hence, there 
are two possibilities of mixing angles both of which reproduce the same 
decay widths of $N_1$ and $N_2$. 

For $J^P=1/2^+$ and $3/2^-$, we display the decay widths and branching 
ratios to the $\pi N$ channel of relevant nucleon resonances in 
Table~\ref{tbl:Ndecay}. Using the mixing angle determined from the mass 
spectrum and experimental information of $N^*\to \pi N$ decays, we obtain 
the decay width of the $\Theta^+$ as shown in Table~\ref{tbl:Thetadecay}.
Among the two values, the former corresponds to the same signs of
$g_{N_{\bm{8}}}$ and $g_{\overline{\bm{10}}}$ (phase 1), while the latter to
the opposite signs (phase 2).

%--table----------------------------
\begin{table}[btp]
    \centering
    \caption{Experimental data for the decay of $N^*$ resonances. We denote 
    the total decay width and partial decay width to the $\pi N$ channel as 
    $\Gamma_{\text{tot}}$ and $\Gamma_{\pi N}$, respectively. Values in 
    parenthesis are the central values quoted in PDG$^{28}$.}
    \vspace{0.2cm}
    \begin{tabular}{|ccrr|}
	\hline
	$J^P$ & Resonance & $\Gamma_{\text{tot}}$ [MeV]
	& Fraction ($\Gamma_{\pi N}/\Gamma_{\text{tot}}$)   \\
	\hline
	$1/2^+$ & $N(1440)$ & 250-450 (350) & 60-70 (65) \% \\
	 & $N(1710)$ & 50-250 (100) & 10-20 (15) \% \\ \hline
	$3/2^-$ & $N(1520)$ & 110-135 (120) & 50-60 (55) \%  \\
	 & $N(1700)$ & 50-150 (100) & 5-15 (10) \%  \\
	 \hline
    \end{tabular}
    \label{tbl:Ndecay}
\end{table}
%--table----------------------------

%--table----------------------------
\begin{table}[btp]
    \centering
    \caption{Decay width of $\Theta^+$ determined from the nucleon decays 
    and the mixing angle obtained from the mass spectra. Phase 1 corresponds
    to the same signs of $g_{N_{\bm{8}}}$ and $g_{\overline{\bm{10}}}$,
    while phase 2 corresponds to the opposite signs. All values are listed 
    in MeV.}
    \vspace{0.2cm}
    \begin{tabular}{|ccll|}
	\hline
	$J^P$ & $\theta_N$ & $\Gamma_{\Theta}$ (Phase 1)
	& $\Gamma_{\Theta}$ (Phase 2)  \\
	\hline
	$1/2^+$ & $29^{\circ}$ (Mass) & 29.1 & 103.3  \\
	$3/2^-$ & $33^{\circ}$ (Mass) & \phantom{0}3.1 & \phantom{0}20.0 \\
	\hline
    \end{tabular}
    \label{tbl:Thetadecay}
\end{table}
%--table----------------------------

Let us adopt the narrower results of phase 1. For the $1/2^+$ case, the 
width is about 30 MeV, which exceeds the upper bound of the experimentally 
observed width. In contrast, the $3/2^-$ case predicts much narrower width 
of the order of a few MeV, which is compatible with the experimental upper 
bound of the $\Theta^+$ width. Considering the agreement of mixing angles 
and the relatively small uncertainties in the experimental decay widths, the
results with the $3/2^-$ case are favorable in the present fitting analysis.

%%%%%%%%%%%%%%%%%%%%%%%%%%%%%%%%%%%%%%%%%%%%%%%%%%%%%%%%%%%%%%%%%%%%%%%%
\section{Two-meson coupling and reaction processes}\label{sec:reaction}
%%%%%%%%%%%%%%%%%%%%%%%%%%%%%%%%%%%%%%%%%%%%%%%%%%%%%%%%%%%%%%%%%%%%%%%%

\begin{figure}[tbp]
    \centering
    \includegraphics[width=5cm,clip]{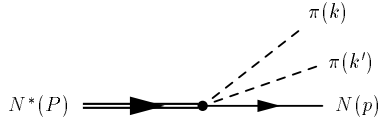}
    \caption{\label{fig:sdecay}
    Feynman diagram for the three-body decay of the $N^*$ resonance.}
\end{figure}

We have performed a phenomenological analysis on the exotic particles using 
flavor SU(3) symmetry. Let us briefly summarize what we have done. It is 
found that the masses of $\Theta(1540)$ and $\Xi_{3/2}(1860)$ are well 
fitted into an antidecuplet ($\bm{\overline{10}}$) representation which 
mixes with an octet ($\bm{8}$), with known baryon resonances of $J^P=1/2^+$ 
or $3/2^-$. Under the representation mixing, the physical nucleon states are
defined as Eq.~\eqref{eq:Nmixing}. Two states $N_1$ and $N_2$ represent 
$N(1440)$ and $N(1710)$ for the $1/2^+$ case, while $N(1520)$ and $N(1700)$ 
for the $3/2^-$ case. The mixing angles $\theta_N$ are determined by 
experimental spectra of known resonances as $\theta_N = 29^{\circ}$ for 
$J^P=1/2^+$ and $\theta_N = 33^{\circ}$ for $J^P=3/2^-$, as shown in
Tables~\ref{tbl:param1} and \ref{tbl:param2}. Using these mixing angles and 
decay widths of nucleon resonances ($\Gamma_{N^*\to\pi N}$), we calculate 
the decay width of $\Theta$ ($\Gamma_{\Theta\to KN}$) through the SU(3) 
relation between the coupling constants. Here we examine the phenomenology 
of three-body decays, as shown in Fig.~\ref{fig:sdecay}.

%--figure---------------------------------
%--figure---------------------------------

In Table~\ref{tbl:exp}, we show the experimental information of the decay 
pattern of the nucleon resonances $N^*\to \pi\pi N$ taken from 
PDG\cite{Eidelman:2004wy}. For convenience, we refer to 
$\pi\pi(I=0, s\text{ wave}) N$ and $\pi\pi(I=1, p\text{ wave}) N$ modes as 
``scalar'' ($s$) and ``vector'' ($v$), respectively. We adopt the total
branching ratio to $\pi\pi N$ channel as the upper limit of the branch for 
$\pi\pi (I=0)N$ state, $BR_{N(1700)\to \pi\pi(I=0)N}<85\text{-}95\% $,
since there is no information for the scalar decay of $N(1700)$.

%--table----------------------------
\begin{table}[btp]
    \centering
    \caption{Experimental information of two-pion decay of nucleon
    resonances. ``Scalar'' represents the mode 
    $\pi\pi(I=0, s\text{ wave}) N$ and ``Vector'' means 
    $\pi\pi(I=1, p\text{ wave}) N$ mode. Values in parenthesis are 
    averaged over the interval quoted in
    PDG$^{28}$.}
    \vspace{0.2cm}
    \begin{tabular}{|c|cccc|}
	\hline
	 $J^P$ & State & $\Gamma_{\text{tot}}$ [MeV]
	 & Scalar [\%]
	 & Vector [\%]	 \\
	\hline
	$1/2^+$ & $N(1440)$
	& 350
	& 5-10(7.5) & $<$8  \\
	& $N(1710)$
	& 100 
	& 10-40(25)  & 5-25(15) \\
	\hline
	$3/2^-$ & $N(1520)$ 
	& 120
	& 10-40(25)  & 15-25(20) \\
	& $N(1700)$ 
	& 100 
	& $<85$-$95$ 
	& $<$35 \\
	\hline
    \end{tabular}
    \label{tbl:exp}
\end{table}
%--table----------------------------

%%%%%%%%%%%%%%%%%%%%%%%%%%%%%%%%%%%%
\subsection{Effective interaction Lagrangians}\label{sec:ReacLag}

Here we write down the effective Lagrangians that account for the relevant
interactions in the present analysis. We need two steps, namely, the 
extraction of the $\tenbar$ component from the $N^*\to \pi\pi N$ decay and 
the extrapolation of that term to the $\Theta \pi K N$ channel. Lagrangians 
for nucleons will be used for the former purpose; the Lagrangians for the 
antidecuplet will tell us the SU(3) relation among different channels in the
multiplet.

Using the partial decay widths of the two nucleon resonances 
$\Gamma^{s,v}_{i}$, we determine the absolute values of the coupling 
constants $|g_{i}^{s,v}|$, where superscripts $s$ and $v$ stand for the 
scalar- and vector-type correlations of two mesons. From them, we can 
obtain the antidecuplet and octet components of the $N^*\pi\pi N$ coupling 
constants as
\begin{equation}
    \begin{split}
	g^{s,v}(\overline{\bm{10}})
	&=-|g_1^{s,v}|\sin\theta_N\pm |g_2^{s,v}|\cos\theta_N ,  \\
	g^{s,v}(\bm{8})
	&=|g_1^{s,v}|\cos\theta_N\pm |g_2^{s,v}|\sin\theta_N ,
    \end{split}
    \label{eq:couplingmix}
\end{equation}
based on Eq.~\eqref{eq:Nmixing}. Here we use $\theta_N$ obtained from the 
mass spectra. We include experimental uncertainties of the $N^*$ decays, 
which will be reflected in uncertainties of $g^{s,v}(\tenbar)$.

To write down the interaction Lagrangians, we adopt the following 
conventions for the fields (nucleons $N$, nucleon resonances $N^*$, pions 
$\bm{\pi}$, octet meson $\phi$, octet baryon $B$ and antidecuplet baryon 
$P$) :
\begin{equation}
    N=
    \begin{pmatrix}
	p \\
	n
    \end{pmatrix},
    \quad
    N^*_i=
    \begin{pmatrix}
	p^*_i \\
	n^*_i
    \end{pmatrix},
    \quad
    \bm{\pi}=
    \begin{pmatrix}
	\pi^0 & \sqrt{2}\pi^+ \\
	\sqrt{2}\pi^- & -\pi^0
    \end{pmatrix}
    \nonumber ,
\end{equation}
\begin{align}
    \phi=&
    \begin{pmatrix}
	\frac{1}{\sqrt{2}}\pi^{0}+\frac{1}{\sqrt{6}}\eta & 
	\pi^{+} & K^{+} \\ 
	\pi^{-} & -\frac{1}{\sqrt{2}}\pi^{0}
	+\frac{1}{\sqrt{6}}\eta & K^{0} \\ 
	K^{-} & \overline{K}^{0} & -\frac{2}{\sqrt{6}}\eta
    \end{pmatrix} ,
    \nonumber
    \\
    B=&
    \begin{pmatrix}
	\frac{1}{\sqrt{2}}\Sigma^{0}+\frac{1}{\sqrt{6}}\Lambda & 
	\Sigma^{+} & p \\ 
	\Sigma^{-} & -\frac{1}{\sqrt{2}}\Sigma^{0}
	+\frac{1}{\sqrt{6}}\Lambda & n \\ 
	\Xi^{-} & \Xi^{0} & -\frac{2}{\sqrt{6}}\Lambda
    \end{pmatrix} ,
    \nonumber
\end{align}
\begin{align}
    P^{333} &= \sqrt{6}\Theta^{+}_{\overline{10}}  , 
    \quad
    P^{133} = \sqrt{2} \, N^0_{\overline{10}}  ,  
    \quad
    P^{233} = -\sqrt{2} \, N^+_{\overline{10}}  ,\nonumber \\*
    P^{113} &= \sqrt{2} \, \Sigma^{-}_{\overline{10}} ,
    \quad
    P^{123} = -\Sigma^{0}_{\overline{10}} ,
    \quad
    P^{223} = -\sqrt{2} \, \Sigma^{+}_{\overline{10}}  , 
    \nonumber  \\*
    P^{111} &=  \sqrt{6}\Xi^{--}_{\overline{10}} , 
    \quad
    P^{112} = -\sqrt{2}  \, \Xi_{\overline{10}}^{-}  ,
    \quad
    P^{122} = \sqrt{2} \, \Xi_{\overline{10}}^{0} , \nonumber \\*
    P^{222} &= - \sqrt{6}\Xi_{\overline{10}}^{+}  .\nonumber
\end{align}

The interaction Lagrangians for nucleons with $J^P=1/2^+$ are
\begin{align}
    \mathcal{L}^{s}_i
    =&\frac{g^{s}_i}{2\sqrt{2}f}
    \overline{N}^*_i\bm{\pi}\cdot \bm{
    \pi} N+ \text{h.c.}
    \nonumber \\
    \mathcal{L}^{v}_i
    =&i\frac{g^{v}_i}{4\sqrt{2}f^2}
    \overline{N}^*_i
    (\bm{\pi}\cdot\Slash{\partial}\bm{\pi}
    -\Slash{\partial}\bm{\pi}\cdot\bm{\pi}) 
    N+ \text{h.c.} ,
    \nonumber
\end{align}
where $f=93$ MeV is the pion decay constant, $g^{s,v}_i$ are dimensionless 
coupling constants, and h.c. stands for the hermitian conjugate. Subscript 
$i=1,2$ denotes the two nucleons $N(1440)$ and $N(1710)$, respectively.
The interaction Lagrangian for the antidecuplet can be written as
\begin{align}
    \mathcal{L}^{s}_{1/2^+}
    =&\frac{g^{s}_{1/2^+}}{2f}
    \overline{P}_{ijk}\epsilon^{lmk}
    \phi_{l}{}^{a}\phi_{a}{}^{i} B_{m}{}^{j}
    + \text{h.c.} ,
    \nonumber \\
    \mathcal{L}^{v}_{1/2^+}
    =&i\frac{g^{v}_{1/2^+}}{4f^2}
    \overline{P}_{ijk}\epsilon^{lmk}\gamma^{\mu}
    (\partial_{\mu}\phi_{l}{}^{a} \phi_{a}{}^{i}
    - \phi_{l}{}^{a}\partial_{\mu} \phi_{a}{}^{i})B_{m}{}^{j}
    + \text{h.c.}
    \nonumber
\end{align}
with the coupling constants for the antidecuplet baryon 
$g^{s}_{1/2^+}$, $g^{v}_{1/2^+}$.

In the same way, the Lagrangians for the $J^P=3/2^-$ case can be written.
We express the spin $3/2$ baryons as Rarita-Schwinger fields 
$B^{\mu}$\cite{Rarita:1941mf}. The effective Lagrangians for nucleons are 
given by
\begin{align}
    \mathcal{L}^{s}_{i}
    &=i\frac{g^{s}_{i}}{4\sqrt{2}f^2}
    \overline{N}_i^{*\mu}\partial_{\mu}(\bm{\pi}\cdot \bm{
    \pi}) N+ \text{h.c.} \nonumber \\
    \mathcal{L}^{v}_{i}
    &=i\frac{g^{v}_{i}}{4\sqrt{2}f^2}
    \overline{N}_{i}^{*\mu}(\bm{\pi}\cdot \partial_{\mu}
    \bm{\pi}-\partial_{\mu}\bm{\pi}\cdot \bm{\pi}) 
    N+ \text{h.c.} 
    \nonumber
\end{align}
Here $i=1,2$ denotes the two nucleons $N(1520)$ and $N(1700)$, respectively.
The Lagrangians for the antidecuplet are constructed as a straightforward 
extension of those for $1/2^+$ case:
\begin{align}
    \mathcal{L}^{s}_{3/2^-}
    &=i\frac{g^{s}_{3/2^-}}{4f^2}
    \overline{P}_{ijk}^{\mu}\epsilon^{lmk}
    \partial_{\mu}(\phi_{l}{}^{a}\phi_{a}{}^{i})
    B_{m}{}^{j}+ \text{h.c.}
    \nonumber \\
    \mathcal{L}^{v}_{3/2^-}
    &=i\frac{g^{v}_{3/2^-}}{4f^2}
    \overline{P}_{ijk}^{\mu}\epsilon^{lmk}
    (\partial_{\mu}\phi_{l}{}^{a} \phi_{a}{}^{i}
    - \phi_{l}{}^{a}\partial_{\mu} \phi_{a}{}^{i})B_{m}{}^{j}
    + \text{h.c.}
    \nonumber
\end{align}

%%%%%%%%%%%%%%%%%%%%%%%%%%%%%%%%%%%%
\subsection{Evaluation of the coupling constants}\label{sec:coupling}

To study the coupling constants, let us start with the decay width of 
a resonance into two mesons and one baryon, which is given by
\begin{align}
    \Gamma_{N\pi\pi}
    =&\frac{M}{16\pi^3}
    \int_{\omega_{\text{min}}}^{\omega_{\text{max}}} d\omega 
    \int_{\omega^{\prime}_{\text{min}}}^{\omega^{\prime}_{\text{max}}} 
    d\omega^{\prime}
    \overline{\Sigma}\Sigma|t(\omega,\omega^{\prime},a)|^2
    \Theta(1-a^2),
    \nonumber
\end{align}
with
\begin{equation}
    \begin{split}
	\omega_{\text{min}} 
	=&\omega^{\prime}_{\text{min}} = m , \quad
	\omega_{\text{max}}
	= \omega^{\prime}_{\text{max}}=\frac{M_R^2-M^2-2Mm}{2M_R} , \\
	a
	=&\cos\theta =  \frac{(M_R-\omega-\omega^{\prime})^2
	-M^2-|\bm{k}|^2-|\bm{k}^{\prime}|^2}{2|\bm{k}||\bm{k}^{\prime}|},
    \end{split}
    \nonumber
\end{equation}
where we assign the momentum variables $P=(M_R,\bm{0})$, 
$k=(\omega,\bm{k})$, $k^{\prime}=(\omega^{\prime},\bm{k}^{\prime})$, and
$p=(E,\bm{p})$ as in Fig.~\ref{fig:sdecay}. $M_R$, $M$, and $m$ are the 
masses of the resonance, baryon, and mesons, respectively, and $\theta$
is the angle between the momenta $\bm{k}$ and $\bm{k}^{\prime}$. The
on-shell energies of particles are given by $\omega=\sqrt{m^2+\bm{k}^2}$, 
$\omega^{\prime}=\sqrt{m^2+(\bm{k}^{\prime})^2}$, and
$E=\sqrt{M^2+\bm{p}^2}$; $\Theta$ denotes the step function; and 
$\overline{\Sigma}\Sigma$ stands for the spin sum of the fermion states.

\begin{figure}[tbp]
    \centering
    \includegraphics[width=5cm,clip]{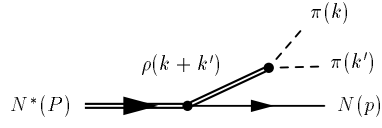}
    \caption{\label{fig:vdecay}
    Three-body decay of the $N^*$ resonance
    with insertion of the vector meson propagator.}
\end{figure}

In the following, we evaluate the squared amplitude 
$\overline{\Sigma}\Sigma|t(\omega,\omega^{\prime},\cos\theta)|^2$
for the $N^*\to \pi\pi N$ decay in the nonrelativistic approximation.
For $J^P=1/2^+$,
\begin{align}
    \overline{\Sigma}\Sigma|t_{1/2^+}^{s}|^2
    =& 3\left(\frac{g^{s}_{1/2^+}}{2f}\right)^2\frac{E+M}{2M}.
    \nonumber \\
    \overline{\Sigma}\Sigma|t^v_{1/2^+}|^2 
    =&
    6\left(\frac{g^{v}_{1/2^+}}{4f^2}\right)^2
    \frac{1}{2M}\Bigl\{
    (E+M)(\omega-\omega^{\prime})^2 
    +2(|\bm{k}|^2-|\bm{k}^{\prime}|^2)(\omega-\omega^{\prime})
    \nonumber \\
    &+(E-M)(\bm{k}-\bm{k}^{\prime})^2
    \Bigr\} 
    \times \left|\frac{-m_{\rho}^2}{s^{\prime}-m_{\rho}^2
    +im_{\rho}\Gamma(s^{\prime})}\right|^2 .
    \nonumber
\end{align}
For the vector-type coupling, we have inserted the vector meson propagator 
to account for the $\rho$ meson correlation\cite{Hosaka:2004mv}, as shown 
in Fig.~\ref{fig:vdecay}. $m_{\rho}$ is the mass of $\rho$ meson, 
$s^{\prime}=(k+k^{\prime})^2$. $\Gamma(s^{\prime})$ is the energy-dependent 
width given by
\begin{equation}
    \Gamma(s^{\prime})
    =\Gamma_{\rho} \times \left(
    \frac{p_{\text{cm}}(s^{\prime})}{p_{\text{cm}}(m_{\rho}^2)}
    \right)^3 ,
    \nonumber
\end{equation}
where $p_{\text{cm}}(s^{\prime})$ is the relative three-momentum of the 
final two particles in the $\rho$ rest frame. For $J^P=3/2^-$, we have
\begin{align}
    \overline{\Sigma}\Sigma|t^{s}_{3/2^-}|^2
    =&\left(\frac{g_{3/2^-}^{s}}{4f^2}\right)^2
    (\bm{k}+\bm{k}^{\prime})^2\frac{E+M}{2M}
    \nonumber ,\\
    \overline{\Sigma}\Sigma|t^{v}_{3/2^-}|^2
    =&
    2
    \left(\frac{g_{3/2^-}^{v}}{4f^2}\right)^2
    (\bm{k}-\bm{k}^{\prime})^2\frac{E+M}{2M}
    \left|\frac{-m_{\rho}^2}{s^{\prime}-m_{\rho}^2
    +im_{\rho}\Gamma(s^{\prime})}\right|^2 .
    \nonumber 
\end{align}

%--figure---------------------------------
%--figure---------------------------------

\begin{figure}[tbp]
    \centering
    \includegraphics[width=5.7cm,clip]{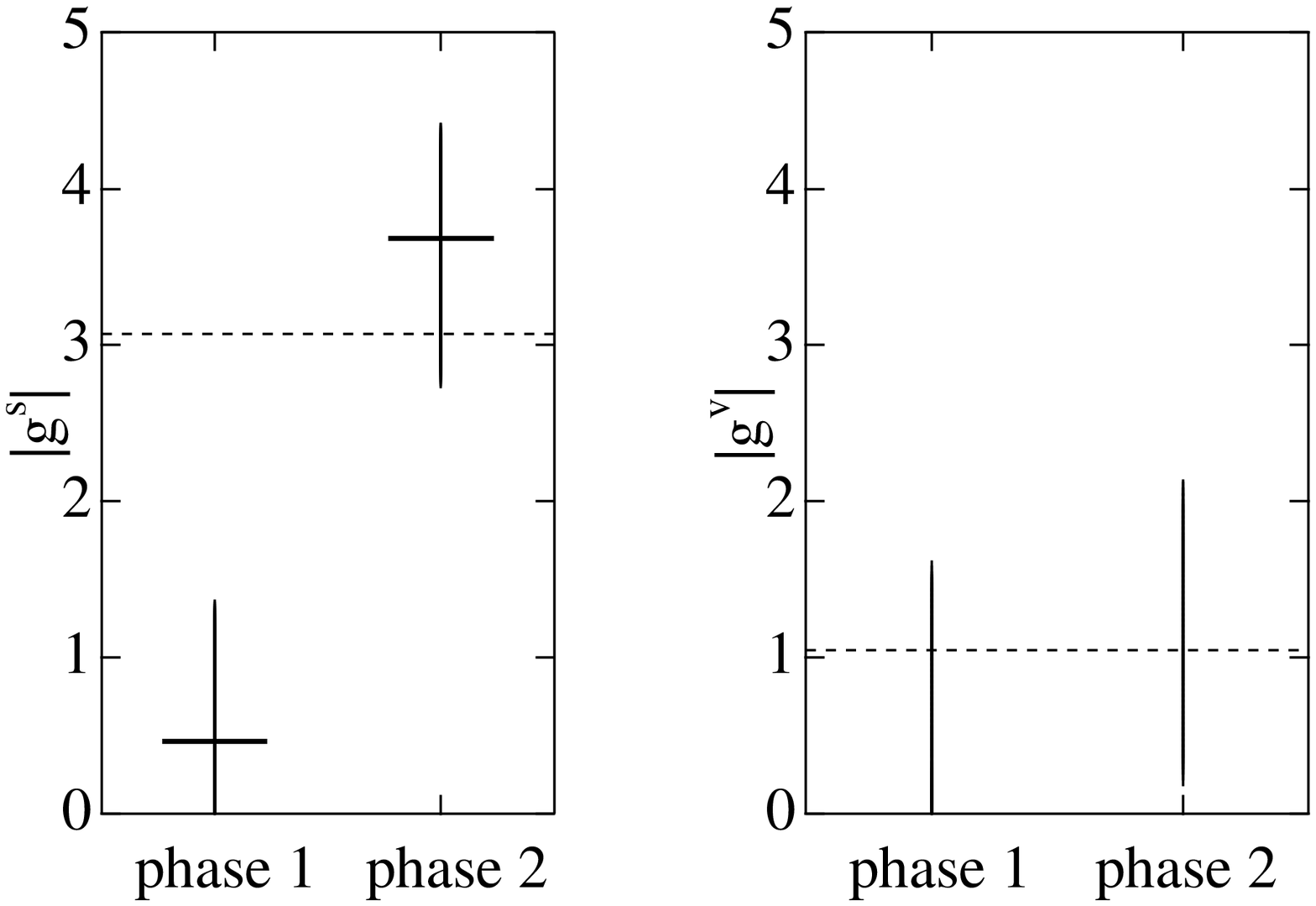}
    \includegraphics[width=5.7cm,clip]{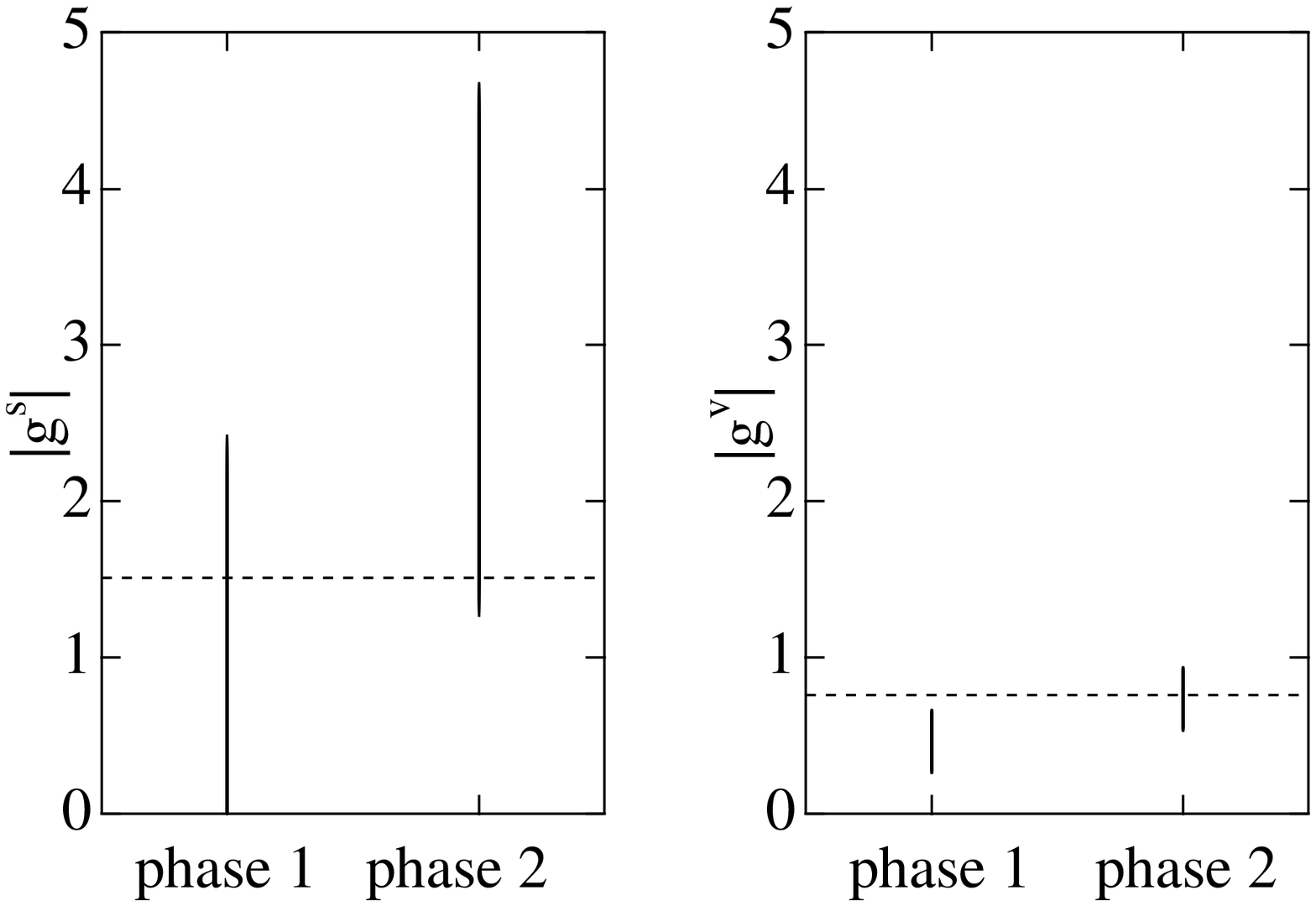}
    \caption{\label{fig:coupling1o2}
    Numerical results for the coupling constants with $J^P=1/2^+$ (Left two 
    panels) and with $3/2^-$ (Right two panels). The two choices of the 
    relative phase between coupling constants are marked as ``phase 1'' and 
    ``phase 2''. Allowed regions of the coupling constants are shown by the 
    vertical bar. Horizontal bars represent the results obtained with the 
    averaged values, which are absent for the vector case. Horizontal dashed
    lines show the upper limits of the coupling constants derived from the 
    self-energy $|\re\Sigma|<200$ MeV.}
\end{figure}

Now we evaluate the coupling constants numerically. Using the averaged
values in Table~\ref{tbl:exp}, we obtain the coupling constants $g^s_i$ and 
$g^v_i$ for these channels. By substituting them into 
Eq.~\eqref{eq:couplingmix} (but suppressing the label $\overline{\bm{10}}$ 
for simplicity), the antidecuplet components are extracted as
\begin{align}
    |g^{s}_{1/2^+}|
    &= 0.47 ,\quad 3.68 ,
    \label{eq:1o2res}
\end{align}
where two values correspond to the results with different relative phases 
between the two coupling constants. When we take into account the 
experimental uncertainties in branching ratio, the antidecuplet components 
can vary within the following ranges:
\begin{align*}
    0
    &< |g^{s}_{1/2^+}| < 1.37,
    & 0
    &< |g^{v}_{1/2^+}|< 2.14,
    \\
    2.72
    &< |g^{s}_{1/2^+}| < 4.42,
\end{align*}
including the two cases of different phase. These uncertainties are also 
shown by the vertical bars in Fig.~\ref{fig:coupling1o2}, with the 
horizontal bars being the result with the averaged value in
Eq.~\eqref{eq:1o2res}.

%--figure---------------------------------
%--figure---------------------------------

Let us consider phenomenological implications of this result. In Ref.~13,
we evaluated the self-energy of the baryon antidecuplet due to two-meson 
coupling. The coupling constants was derived by assuming that the $\Theta^+$
belongs to a pure antidecuplet together with $N(1710)$, where we determined 
$|g^{s}_{1/2^+}|=1.88$ and $|g^{v}_{1/2^+}|=0.315$. In the calculation of 
the self-energy of $\Theta^+$, the effect of the mixing only changes the 
coupling constants, by neglecting the small contribution from 
$\mathcal{L}^{27}$. In this case, the $\Theta^+$ self-energy with the new 
coupling constants can be obtained by simple rescaling. Using the values 
of Eq.~\eqref{eq:1o2res}, we estimate
\begin{align}
    \Sigma^s_{\Theta^+}
    &= -287, \quad -4.7 \text{ MeV} ,
    &0>\Sigma^v_{\Theta^+}
    &> -770\text{ MeV} .
    \nonumber
\end{align}
The sum of these values are the contribution to the self-energy of 
$\Theta^+$ from the two-meson cloud. Naively, we expect that it should be 
of the order of 100 MeV, at most $\sim$20\% of the total 
energy\cite{Hosaka:2004mv}. From this consideration, we adopt the condition 
that the magnitude of one of the contributions should not exceed $200$ MeV:
$|\re\Sigma_{\Theta^+}|<200$. This condition is satisfied when
\begin{equation}
    |g^{s}_{1/2^+}|<3.07, \quad |g^{v}_{1/2^+}|<1.05.
    \label{eq:slimit}
\end{equation}
Therefore, we can exclude the choice of ``phase 2'' in left two panels in
Fig.~\ref{fig:coupling1o2}. The limit for $|g^{v}_{1/2^+}|$ is compatible 
with Eq.~\eqref{eq:1o2res}, although the self-energy argument gives a more 
stringent constraint. These upper limits are also shown in 
Fig.~\ref{fig:coupling1o2} by the dashed lines.

Now we consider the $J^P=3/2^-$ case. Using the central values in 
Table~\ref{tbl:exp}, we obtain the coupling constants $g^s_i$ and $g^v_i$ 
for these channels. In this case, with the same reason as in the vector 
coupling for the $1/2^+$ case, the central value cannot be determined. 
Experimental uncertainties allows the antidecuplet components to vary within
the following ranges:
\begin{align}
    0
    &< |g^{s}_{3/2^-}| < 4.68,
    &0.25
    &< |g^{v}_{3/2^-}|< 0.94,
    \nonumber
\end{align}
including the two cases of different phase. The results are shown by the 
vertical bars in right two panels of Fig.~\ref{fig:coupling1o2}.

We can also estimate the magnitude of the self-energy, by substituting the 
squared amplitudes for $3/2^-$ case in the formulas of the self-energy 
shown in Ref.~13. For instance, we estimate the real part of the self-energy
as $-1518$ MeV for an initial energy of $1540$-$1700$ MeV and a cutoff of 
$700$-$800$ MeV for $|g^{s}_{3/2^-}|=  4.17$. The huge self-energy is due to
the $p$-wave nature of the two-meson coupling. In this way, to have some 
reasonable values for the self-energy $|\re\Sigma_{\Theta^+}|<200$ MeV,
\begin{equation}
    |g^{s}_{3/2^-}|<1.51 , \quad |g^{v}_{3/2^-}|<0.76 .
    \nonumber
\end{equation}
Both upper limits are indicated by horizontal dashed lines in the 
right two panels in Fig.~\ref{fig:coupling1o2}.

%%%%%%%%%%%%%%%%%%%%%%%%%%%%%%%%%%%%
\subsection{Analysis of the meson-induced reactions}\label{sec:Reacreaction}

\begin{figure}[tbp]
    \centering
    \includegraphics[width=7cm,clip]{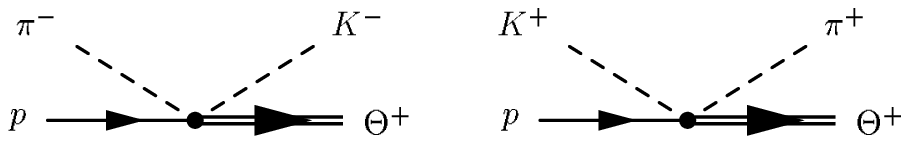}
    \hspace{0.5cm}
    \includegraphics[width=3cm,clip]{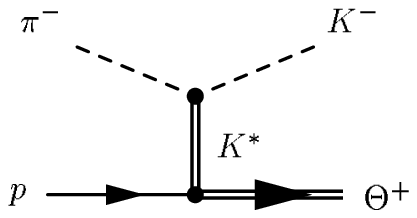}
    \caption{\label{fig:reaction}
    Feynman diagrams for the meson-induced reactions for $\Theta^+$ 
    production. Right : $\piK$ reaction. Center : $\Kpi$ reaction.
    Left : $\piK$ reaction with a vector meson propagator.}
\end{figure}

As an application of effective Lagrangians, we calculate the reaction
processes $\piK$ and $\Kpi$ \textit{via} tree-level diagrams as shown in 
Fig.~\ref{fig:reaction}. These are alternative reactions to, for instance, 
photo-induced reactions, which are useful for further study of the 
$\Theta^+$. The amplitudes for these reactions are given by
\begin{align}
    -it_{1/2^+}^{s}(\piK)
    =&-it_{1/2^+}^{s}(\Kpi)\nonumber \\
    =&i\frac{g_{1/2^+}^{s}}{2f}(-\sqrt{6})
    N_{\Theta^+}N_p , 
    \label{eq:samp1o2reac} \\
    -it_{1/2^+}^{v}(\piK)
    =&it_{1/2^+}^{v}(\Kpi)\nonumber \\
    =& i\frac{g_{1/2^+}^{v}}{4f^2}(-\sqrt{6})
    (2\sqrt{s}-M_{\Theta}-M_p) 
    N_{\Theta^+}N_p
    F(k-k^{\prime}) 
    \label{eq:vamp1o2reac}
\end{align}
for the $1/2^+$ case and by
\begin{align}
    -it_{3/2^-}^{s}(\piK)
    =&-it_{3/2^-}^{s}(\Kpi)\nonumber \\
    =&i\frac{g_{3/2^-}^{s}}{4f^2}(-\sqrt{6})
    (\bm{k}-\bm{k}^{\prime})\cdot \bm{S}^{\dag}
    N_{\Theta^+}N_p , 
    \nonumber \\
    -it_{3/2^-}^{v}(\piK)
    =&it_{3/2^-}^{v}(\Kpi)\nonumber \\
    =& -i\frac{g_{3/2^-}^{v}}{4f^2}(-\sqrt{6})
    (\bm{k}+\bm{k}^{\prime})\cdot \bm{S}^{\dag}
    N_{\Theta^+}N_p
    F(k-k^{\prime})
    \label{eq:vamp3o2reac}
\end{align}
for the $3/2^-$ case, where the normalization factor is 
$N_i=\sqrt{(E_i+M_i)/2M_i}$, $\bm{S}$ is the spin transition operator,
$\sqrt{s}$ is the initial energy, 
and $k$ and $k^{\prime}$ are the momenta of the incoming and outgoing 
mesons, respectively. Here we define the vector meson propagator 
(Fig.~\ref{fig:reaction}, right) as
\begin{equation}
    F(k-k^{\prime})=\frac{-m_{K^*}^2}{(k-k^{\prime})^2-m_{K^*}^2
    +im_{K^*}\Gamma[(k-k^{\prime})^2]},
    \label{eq:formfactor}
\end{equation}
which is included in the vector-type amplitude. In the kinematical region 
in which we are interested, the momentum-dependent decay width of $K^*$, 
$\Gamma[(k-k^{\prime})^2]$ vanishes. 

%--figure---------------------------------
%--figure---------------------------------

Since the two amplitudes must be summed coherently, the squared amplitudes 
are given by
\begin{align}
    \overline{\Sigma}\Sigma|t_{1/2^+}|^2 
    = &
    6\left(\frac{1}{2f}\right)^2 N_{\Theta^+}^2N_p^2
    \Bigl[(g_{1/2^+}^{s})^2 
    \nonumber \\
    & 
    \pm 2g_{1/2^+}^{s}g_{1/2^+}^{v} 
    \frac{2\sqrt{s}-M_{\Theta}-M_p}{2f}F(k-k^{\prime}) \nonumber \\
    & 
    + (g_{1/2^+}^{v})^2
    \frac{(2\sqrt{s}-M_{\Theta}-M_p)^2}{4f^2}F^2(k-k^{\prime}) 
    \Bigr] ,
    \label{eq:ampsqure1o2}  \\
    \overline{\Sigma}\Sigma|t_{3/2^-}|^2  
    = &4
    \left(\frac{1}{4f^2}\right)^2 
    N_{\Theta^+}^2N_p^2
    \Bigl[
    (g_{3/2^-}^{s})^2 (\bm{k}-\bm{k}^{\prime})^2 \nonumber \\
    & 
    \mp 2g_{3/2^-}^{s}g_{3/2^-}^{v}(|\bm{k}|^2-|\bm{k}^{\prime}|^2)
    F(k-k^{\prime}) \nonumber \\
    &+(g_{3/2^-}^{v})^2(\bm{k}+\bm{k}^{\prime})^2F^2(k-k^{\prime})
    \Bigr] ,
    \nonumber
\end{align}
where $\pm$ and $\mp$ signs denote the $\piK$ and $\Kpi$ reactions,
respectively. Notice that the relative phase between the two coupling
constants is important, which affects the interference term of the two 
amplitudes. To determine the phase, we use the experimental information
from $\piK$ reaction at KEK\cite{ImaiMiwa,Miwa:2006if}, where the upper 
limit of the cross section has been extracted to be a few $\mu$b.

The differential cross section for these reactions is given by 
\begin{align}
    \frac{d\sigma}{d\cos\theta} (\sqrt{s},\cos\theta)  
    =&\frac{1}{4\pi s}\frac{|\bm{k}^{\prime}|}{|\bm{k}|}
    M_{p}M_{\Theta}
    \frac{1}{2}
    \overline{\Sigma}\Sigma|t(\sqrt{s},\cos\theta)|^2  ,
    \nonumber
\end{align}
which is evaluated in the center-of-mass frame. The total cross section
can be obtained by integrating this over $\cos\theta$.

%%%%%%%%%%%%%%%%%%%%%%%%%%%%%%%%%%%%
\subsection{Qualitative features of the cross sections}

\begin{figure}[tbp]
    \centering
    \includegraphics[width=5.8cm,clip]{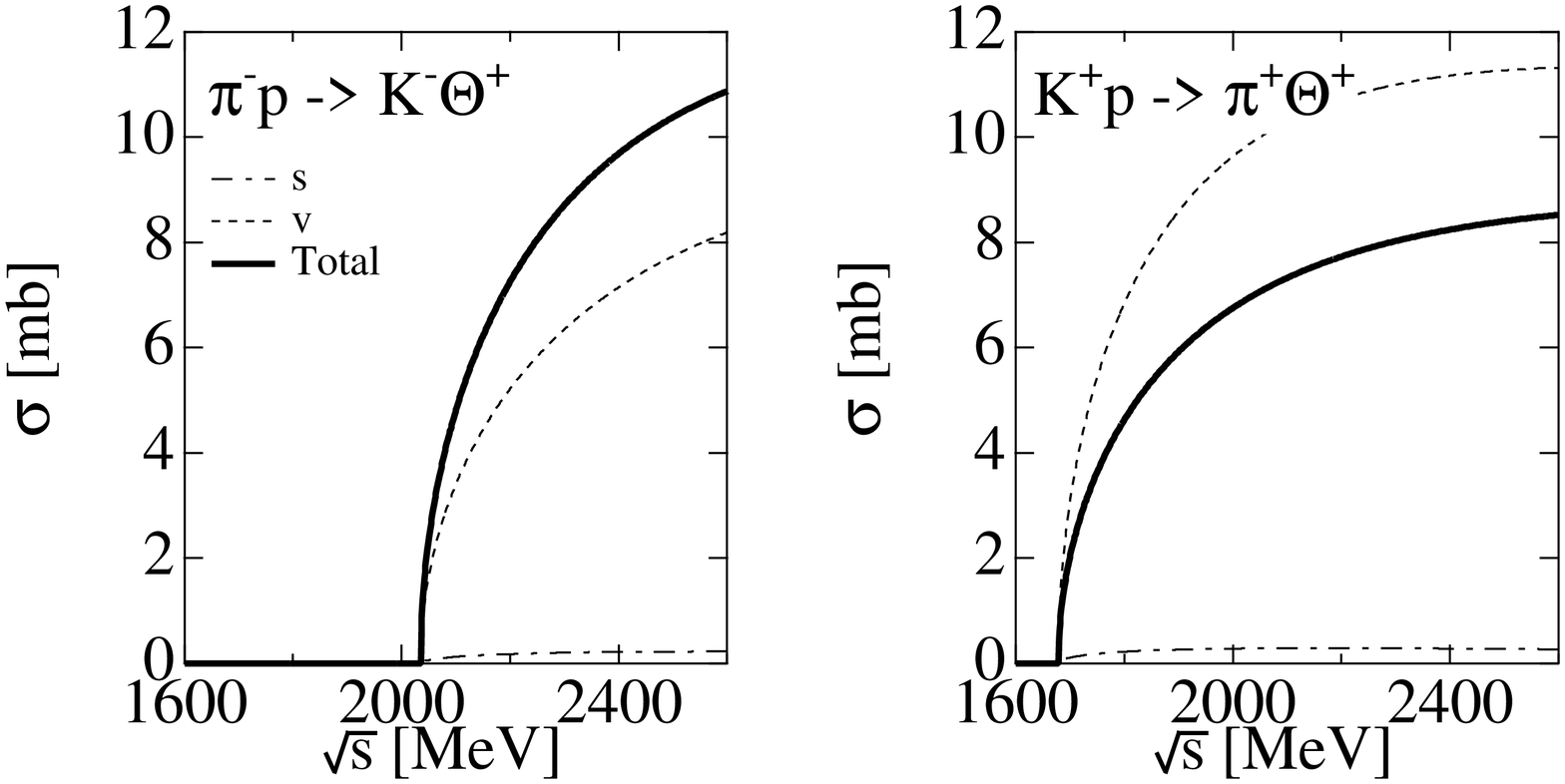}
    \includegraphics[width=5.8cm,clip]{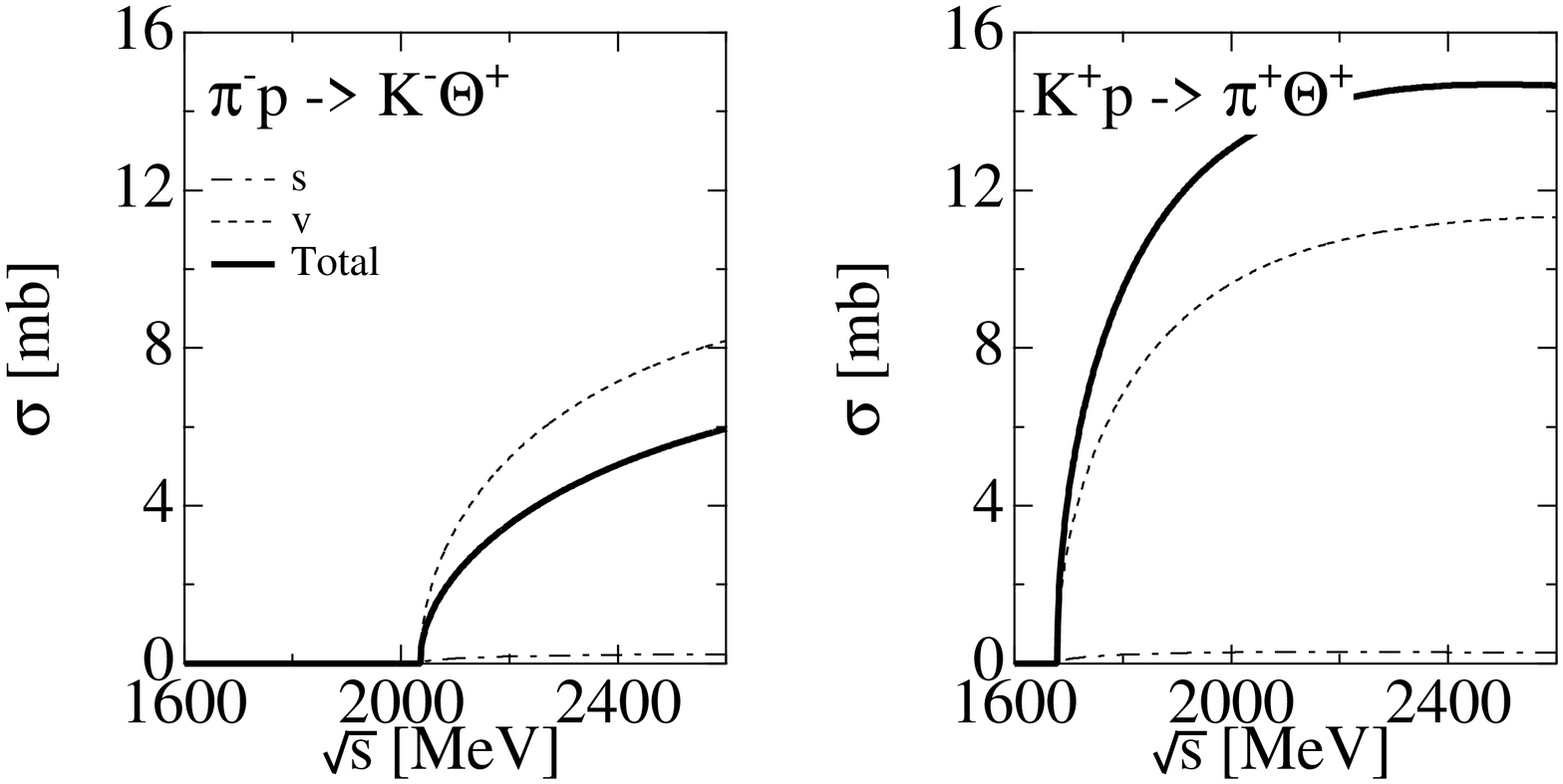}
    \caption{\label{fig:1o2}
    Total cross sections for the $J^P=1/2^+$ case with $g^s=0.47$ and 
    $g^v=0.47$ (Left two panels) and with $g^v=-0.47$ (Right two panels). 
    The thick line shows the result with full amplitude. Dash-dotted and 
    dashed lines are the contributions from $s$ and $v$ terms, 
    respectively.}
\end{figure}

Now let us calculate the cross section using the coupling constants obtained
previously. Here we focus on the qualitative difference between 
$J^P=1/2^+$ and $3/2^-$ cases. We first calculate for the $1/2^+$ case, with
coupling constants
\begin{equation}
    g^s_{1/2^+} = 0.47,\quad g^v_{1/2^+} = 0.47,
    \label{eq:1o2_1}
\end{equation}
where $g^s_{1/2^+}$ is one of the solutions that satisfies the 
condition~\eqref{eq:slimit}. Since the result for $g^v_{1/2^+}$ spreads 
over a wide range, we choose $g^v_{1/2^+}=g^s_{1/2^+}$, which is well within
the allowed region determined from the self-energy. The result is shown in 
the left two panels in Fig.~\ref{fig:1o2}, with individual contributions 
from $s$ and $v$ terms. As we see, the use of the same coupling constant for
both terms results in the dominance of the vector term. However, there is a 
sizable interference effect between $s$ and $v$ terms, although the 
contribution from the $s$ term itself is small. The two amplitudes interfere
constructively for the $\piK$ channel, whereas in the $\Kpi$ case they 
destructively interfere.

%--figure---------------------------------
%--figure---------------------------------

As already mentioned, the relative phase of the two coupling constants 
is not determined. If we change the sign,
\begin{equation}
    g^s_{1/2^+} = 0.47,\quad g^v_{1/2^+} = -0.47,
    \label{eq:1o2_2}
\end{equation}
then the results change as in the right two panels in Fig.~\ref{fig:1o2}, 
where constructive and destructive interference appears in an opposite 
manner. 

There is an experimental result from KEK\cite{Miwa:2006if} that the 
cross section of $\piK$ was found to be very small, of the order of a few 
$\mu$b. This is an order smaller than the typical background processes of 
this energy, which are of the order of 30 $\mu$b\cite{Miwa:2006if}. At this
stage, we do not want to calculate the cross section quantitatively, but the
experimental result suggests that the choice of Eq.~\eqref{eq:1o2_2} should 
be plausible, for the small cross section of the $\piK$ reaction. In this 
case, the cross section for $\Kpi$ becomes large.

\begin{figure}[tbp]
    \centering
    \includegraphics[width=5.8cm,clip]{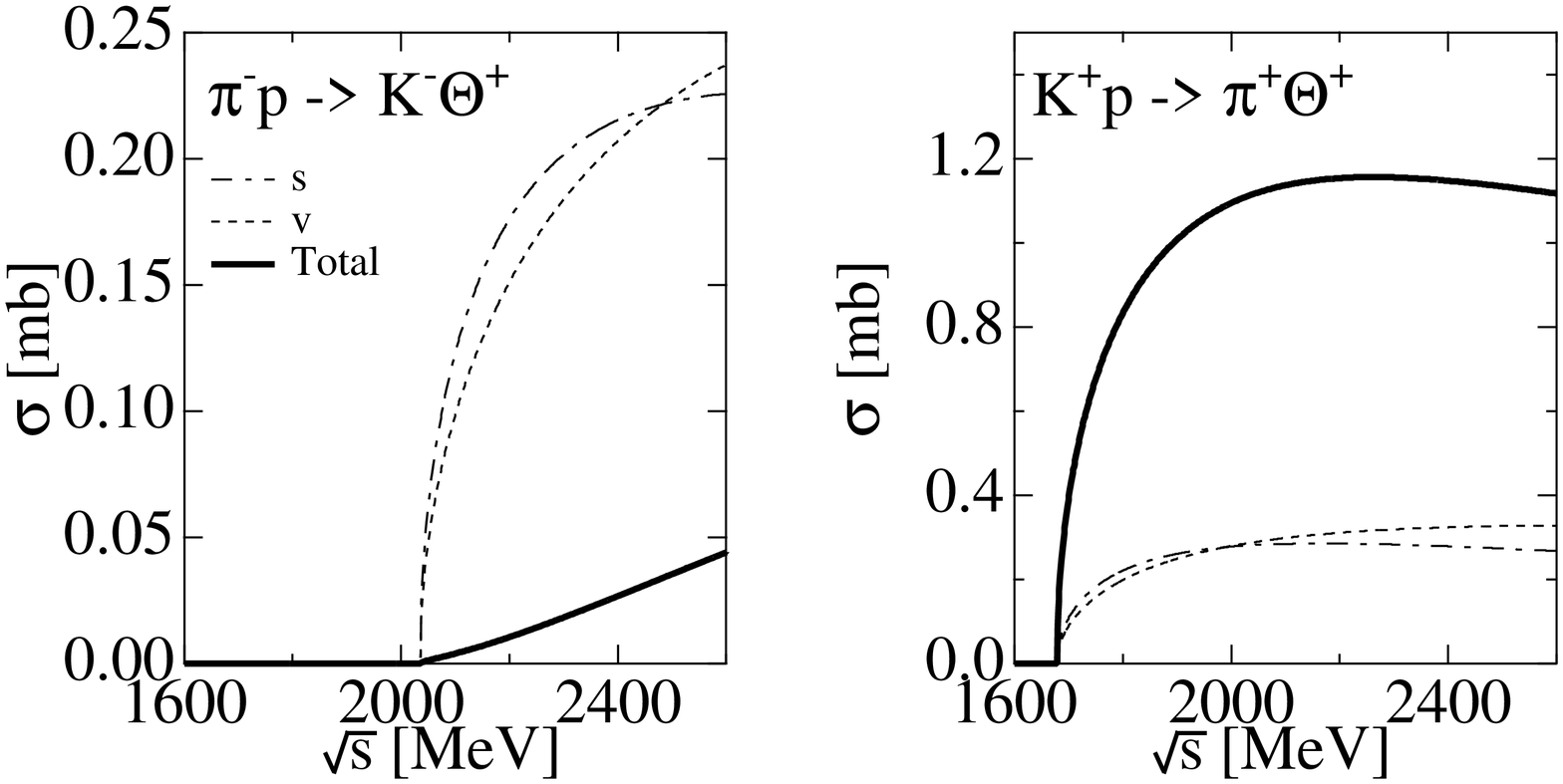}
    \includegraphics[width=5.8cm,clip]{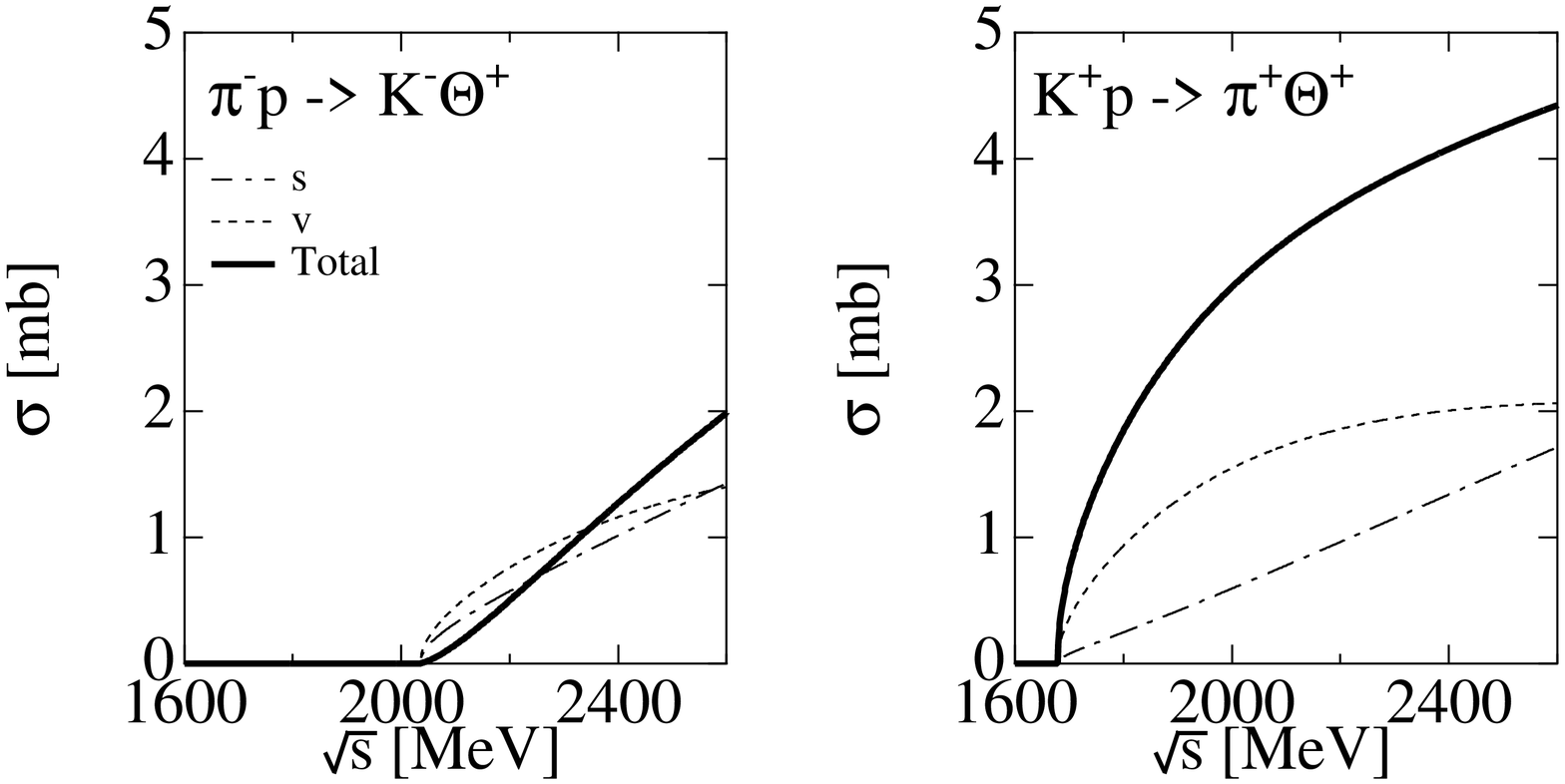}
    \caption{\label{fig:1o2_3}
    Total cross sections for the $J^P=1/2^+$ case with $g^s=0.47$ and 
    $g^v=-0.08$ (Left two panels), and for the $J^P=3/2^-$ case with 
    $g^s=0.2$ and $g^v=0.4$ (Right two panels), when the most destructive 
    interference for $\piK$ takes place. Note that the vertical scale is 
    different in the two panels. The thick line shows the result with full 
    amplitude. Dash-dotted and dashed lines are the contributions from $s$ 
    and $v$ terms, respectively.}
\end{figure}

As a trial, let us search for the set of coupling constants with which the 
most destructive interference takes place in $\piK$, by changing
$g^v_{1/2^+}$ within the allowed region. This means that the 
difference between cross sections of $\piK$ and $\Kpi$ is maximal. Then we 
find
\begin{equation}
    g^s_{1/2^+} = 0.47,\quad g^v_{1/2^+} = -0.08.
    \label{eq:1o2_3}
\end{equation}
The result is shown in left two panels in Fig.~\ref{fig:1o2_3}. A huge 
difference between $\piK$ and $\Kpi$ can be seen. In this case, we observe 
the ratio of cross sections
\begin{equation}
    \frac{\sigma(\Kpi)}{\sigma(\piK)}
    \sim 50 ,
    \nonumber
\end{equation}
where we estimated the cross section $\sigma$ as the average of the cross
section shown in the figures (from threshold to 2.6 GeV). Notice that the
ratio of the coupling constants $g^s_{1/2^+}/g^v_{1/2^+} \sim -5.9$ is
relevant for the interference effect. It is possible to scale both coupling 
constants within experimental uncertainties. This does not change the ratio 
of cross sections, but it does change the absolute values.

%--figure---------------------------------
%--figure---------------------------------

Next we examine the case with $J^P=3/2^-$. Again, we observe constructive 
and destructive interferences, depending on the relative sign of the two 
amplitudes. The interference effect is prominent around the energy region 
close to the threshold but is not very strong in the higher energy region, 
compared with $1/2^+$ case.
We search for the coupling constants with which the most destructive 
interference takes place for $\piK$. We find that destructive interference 
is maximized when the ratio of the coupling constants is  $g^s_{3/2^-}
/g^v_{3/2^-} \sim 0.5$. Taking, for instance, the values
\begin{equation}
    g^s_{3/2^-} = 0.2,\quad g^v_{3/2^-} = 0.4,
    \label{eq:3o2_3}
\end{equation}
which are within the experimental bounds, we obtain the results shown in 
right two panels in Fig.~\ref{fig:1o2_3}. In contrast to the $J^P=1/2^+$ 
case, here the ratio of cross section is not very large:
\begin{equation}
    \frac{\sigma(\Kpi)}{\sigma(\piK)}
    \sim 3.3 .
    \nonumber
\end{equation}
The high-energy behavior in this case is understood from the $p$-wave 
nature of the coupling.

%%%%%%%%%%%%%%%%%%%%%%%%%%%%%%%%%%%%
\subsection{Quantitative analysis}

Here we consider the reaction mechanism in detail to provide a more 
quantitative result. First we introduce a hadronic form factor at the 
vertices, which accounts for the energy dependence of the coupling 
constants. Physically, it is understood as the reflection of the finite 
size of the hadrons. In practice, however, the introduction of the form
factor has some ambiguities in its form and the cutoff 
parameters\cite{Nam:2003uf}.

\begin{figure}[tbp]
    \centering
    \includegraphics[width=5.8cm,clip]{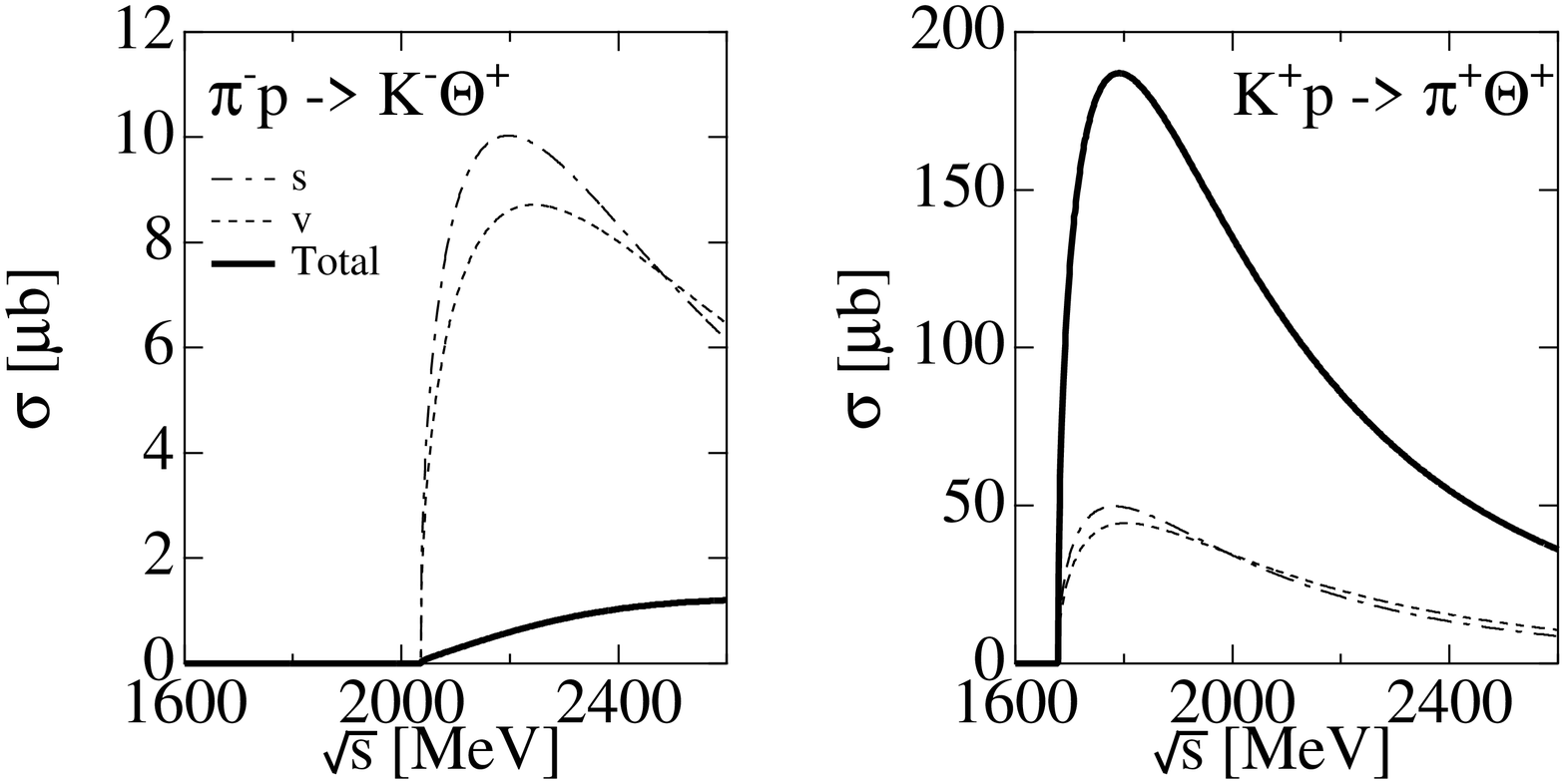}
    \includegraphics[width=5.8cm,clip]{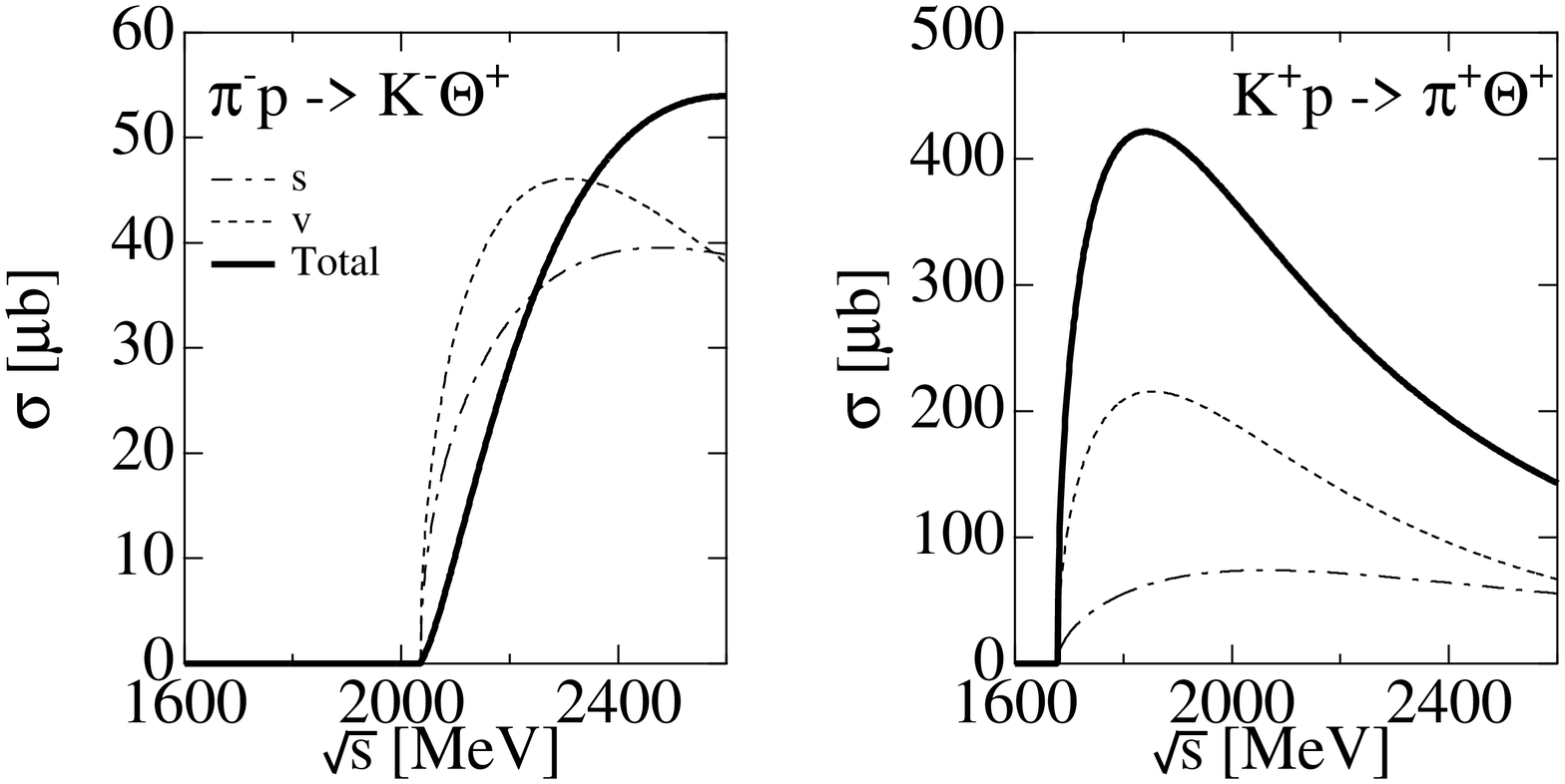}
    \caption{\label{fig:1o2_hFF}
    Total cross sections for the $J^P=1/2^+$ case with $g^s=0.47$ and 
    $g^v=-0.08$ (Left two panels), and for the $J^P=3/2^-$ case with 
    $g^s=0.2$ and $g^v=0.4$ including a hadronic form 
    factor~\eqref{eq:monopoleFF}. 
    The thick line shows the result with full amplitude. Dash-dotted and 
    dashed lines are the contributions from $s$ and $v$ terms, 
    respectively.}
\end{figure}

In Ref.~23, the $\piK$ reaction is studied with a three-dimensional 
monopole-type form factor
\begin{equation}
    F(\sqrt{s})=\frac{\Lambda^2}{\Lambda^2+\bm{q}^2},
    \label{eq:monopoleFF}
\end{equation}
where $\bm{q}^2=\lambda(s,M_N^2,m_{\text{in}}^2)/4s$ with $m_{\text{in}}$ 
being the mass of the incoming meson and $\Lambda=0.5$ GeV. Here we adopt 
this form factor and apply it to the present process. We obtain the results 
for $J^P=1/2^+$ and for $J^P=3/2^-$ in left two panels and right two 
panels in Fig.~\ref{fig:1o2_hFF}, with the coupling constants given in 
Eqs.~\eqref{eq:1o2_3} and \eqref{eq:3o2_3}. 

%--figure---------------------------------
%--figure---------------------------------

We observe that the cross section is suppressed down to $\sim 1 \mu$b 
for the $\piK$ reaction in the $1/2^+$ case. However, this is also a 
consequence of our choice of small coupling constants. Indeed, with these
coupling constants, the self-energy of $\Theta^+$ becomes
\begin{equation}
    \re\Sigma_{\Theta^+}^{1/2^+}
    = \re\Sigma^s_{\Theta^+} 
    +\re\Sigma^v_{\Theta^+}
    \sim -5.3-1.6=-6.9 \text{ MeV},
    \nonumber
\end{equation}
for $p^{0}=1700$ MeV and a cutoff $750$ MeV. This is too small, but as we
mentioned before, we can scale these constants without changing the ratio 
of $\Kpi$ and $\piK$. We would like to search for the coupling constants
which provide a small cross section for $\piK$ reaction compatible with
experiment and a moderate amount of self-energy, which guarantee the
dominance of the two-meson coupling terms compared with the $KN\Theta^+$ 
vertex.

\begin{figure*}[tbp]
    \centering
    \includegraphics[width=5cm,clip]{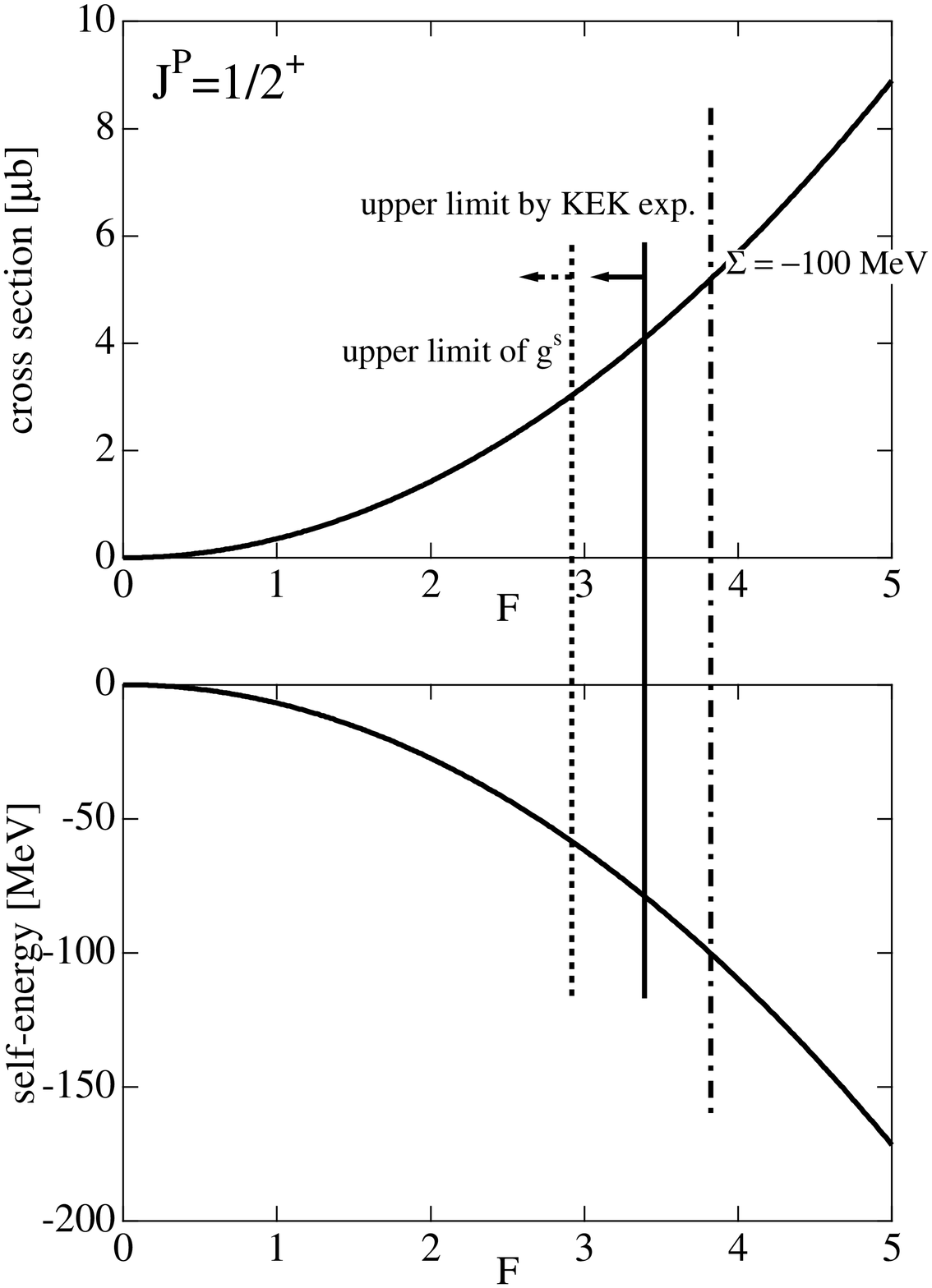}
    \includegraphics[width=5cm,clip]{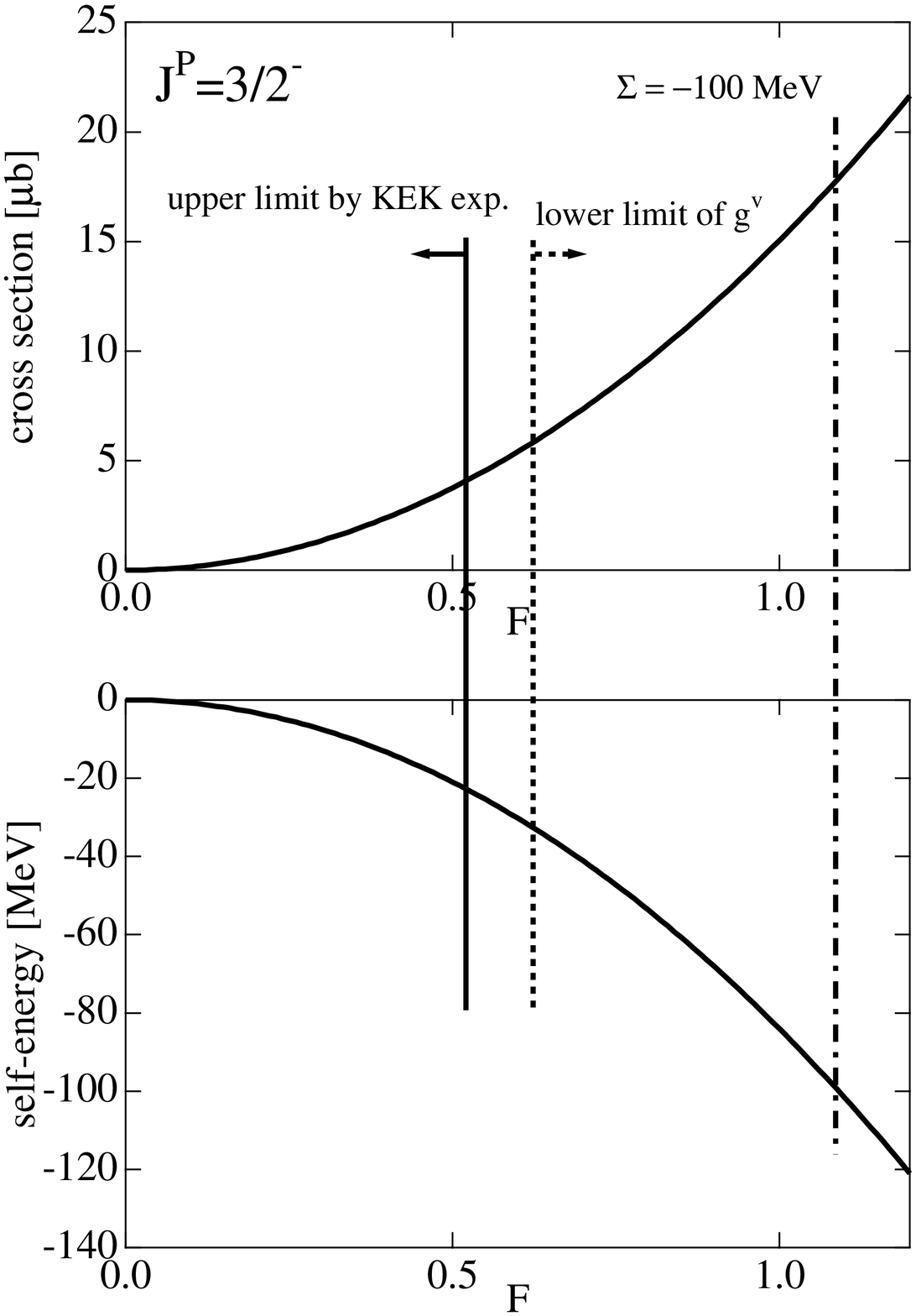}
    \caption{\label{fig:selfcross}
    The total cross section of $\piK$ at $P_{\text{lab}}=1920$ MeV and the 
    real part of the self-energy of $\Theta^+$ as functions of the factor 
    $F$ defined in Eqs.~\eqref{eq:factor} and \eqref{eq:factor3o2} for 
    $J^P=1/2^+$ (left) and $J^P=3/2^-$ (right). Solid, dashed and 
    dash-dotted vertical lines show the upper limit of cross section given 
    by the KEK experiment$^{31,24}$, the limit of coupling
    constant, and the point where $\re\Sigma=-100$ MeV.}
\end{figure*}

In Fig.~\ref{fig:selfcross}, we plot the cross section of $\piK$ reaction
and the self-energy of $\Theta^+$ by fixing the ratio of coupling 
constants. The cross section is the value at $\sqrt{s}=2124$ MeV, which
corresponds to the KEK experiment $P_{\text{lab}}\sim$ 1920 MeV. The 
horizontal axis denotes the factor $F$, which is defined by
\begin{equation}
    g^s_{1/2^+} = F\times 0.47,\quad
    g^v_{1/2^+} = -F\times 0.08 .
    \label{eq:factor}
\end{equation}
We use $F=1$ for the calculation of Fig.~\ref{fig:1o2_hFF}. Both the cross 
section and self-energy are proportional to the square of the coupling 
constant. By vertical bars, we indicate the points where
\begin{itemize}
    \item   cross section $\sim 4.1\mu$b\footnote{Here we use the 
    preliminary value $4.1 \mu$b reported in Ref.~31, which has been later 
    corrected as $3.9 \mu$b~\cite{Miwa:2006if}. Qualitative conclusions 
    remain unchanged for the upper limit of 3.9 $\mu$b.} 
    estimated by KEK experiment\cite{ImaiMiwa,Miwa:2006if}

    \item  upper limit of $g^s_{1/2^+}$

    \item   $\re\Sigma_{\Theta}=-100$ MeV
\end{itemize}

%--figure---------------------------------
%--figure---------------------------------

For the $J^P=3/2^-$ case, with $g^s_{3/2^-}=0.2$ and $g^v_{3/2^-}=0.4$, the 
self-energy of $\Theta^+$ becomes
\begin{equation}
    \re\Sigma_{\Theta^+}^{3/2^-}
    = \re\Sigma^s_{\Theta^+}
    +\re\Sigma^v_{\Theta^+} 
    \sim -4-80=-84 \text{ MeV}.
    \nonumber
\end{equation}
In Fig.~\ref{fig:selfcross}, we plot the cross section of the $\piK$ 
reaction and the self-energy of $\Theta^+$ by fixing the ratio of coupling 
constants. The horizontal axis denotes the factor $F$, which is defined by
\begin{equation}
    g^s_{1/2^+} = F\times 0.2,\quad
    g^v_{1/2^+} = F\times 0.4.
    \label{eq:factor3o2}
\end{equation}
As in the same way, we indicate
\begin{itemize}
    \item   cross section $\sim 4.1\mu$b

    \item  lower limit of $g^v_{3/2^-}$

    \item   $\re\Sigma_{\Theta}=-100$ MeV
\end{itemize}
These results, as well as the corresponding cross section for the $\Kpi$ 
reaction are summarized in Table~\ref{tbl:summary}.

%--table----------------------------
\begin{table}[btp]
    \centering
    \caption{Summary of the coupling constants, cross sections and 
    self-energies. $\sigma_{\pi^-}$ is the total cross section for $\piK$
    are the values at $P_{\text{lab}}=1920$ MeV; $\sigma_{K^+}$ is that 
    for $\Kpi$, which is the upper limit of the cross section at 
    $P_{\text{lab}}=1200$ MeV.}
    \vspace{0.2cm}
    \begin{tabular}{|l|lllll|}
	\hline
	$J^P$ & $g^s$ & $g^v$ & $\sigma_{\pi^-}$ [$\mu$b]
	& $\sigma_{K^+}$ [$\mu$b]
	& $\re\Sigma_{\Theta}$ [MeV]  \\
	\hline
	$1/2^+$  & 1.59 & $-0.27$
	& \phantom{0}4.1 & $<$1928 & \phantom{0}$-78$  \\
	 & 1.37 & $-0.23$
	 & \phantom{0}3.2 & $<$1415 & \phantom{0}$-58$  \\
	 & 1.80 & $-0.31$ & \phantom{0}5.0 & $<$2506 & $-100$  \\
	 \hline
	$3/2^-$ & 0.104 & 0.209 & \phantom{0}4.1 & $<$\phantom{0}113 
	& \phantom{0}$-23$  \\
	& 0.125 & 0.25 & \phantom{0}5.9 & $<$\phantom{0}162 
	& \phantom{0}$-32$  \\
	 & 0.22 & 0.44 & 18 & $<$\phantom{0}502 & $-100$  \\
	 \hline
    \end{tabular}
    \label{tbl:summary}
\end{table}
%--table----------------------------

In Ref.~2, we have calculated the angler dependence of the differential 
cross sections at the energy of KEK experiment: $P_{\text{lab}}\sim 1920$ 
MeV for $\piK$ and $P_{\text{lab}}\sim 1200$ MeV for $\Kpi$. The 
cross sections show forward peaking behavior, although the dependence 
is not very strong.

Finally, we would like to briefly discuss the possible effect from the Born
terms. There are reasons that the Born terms are not important in the 
present reactions. First, the cross sections of the Born terms are 
proportional to the decay width of $\Theta^+$ and therefore suppressed if 
the decay width of the $\Theta^+$
is narrow. Second, in the energy region of $\Theta^+$ production, the energy
denominator of the exchanged nucleon suppresses the contribution, especially
for the $s$-channel term in the $\piK$ reaction. Indeed, as demonstrated 
explicitly in Ref.~2, the effect of Born term is small enough to be 
neglected, for the amplitudes of the two-meson couplings studied here.
However, if the experimental cross section for $\Kpi$ is much smaller than 
we expected, the smaller coupling constants for the two-meson coupling are 
required. In such a case, we should consider the Born terms and interference
effects more seriously.

%%%%%%%%%%%%%%%%%%%%%%%%%%%%%%%%%%%%%%%%%%%%%%%%%%%%%%%%%%%%%%%%%%%%%%
\section{Summary}\label{sec:summary}
%%%%%%%%%%%%%%%%%%%%%%%%%%%%%%%%%%%%%%%%%%%%%%%%%%%%%%%%%%%%%%%%%%%%%%

We have studied masses and decay widths of the baryons belonging to the 
octet ($\bm{8}$) and antidecuplet ($\overline{\bm{10}}$) based on the flavor
SU(3) symmetry. As pointed out 
previously\cite{Cohen:2004gu,Pakvasa:2004pg,Praszalowicz:2004xh}, we 
confirmed again the inconsistency between the mass spectrum and decay widths
of flavor partners in the $\bm{8}$-$\overline{\bm{10}}$ mixing scenario with
$J^P=1/2^+$. However, the assignment of $J^P=3/2^-$ particles in the mixing 
scenario well reproduces the mass spectrum as well as the decay widths of 
$\Theta(1540)$, $N(1520)$, 
and $N(1700)$. Assignment of $3/2^-$ predicts a new $\Sigma$ state at around
1840 MeV, and the nucleon mixing angle is close to the one of ideal mixing. 
The $1/2^-$ assignment is not realistic since the widths are too large for 
$\Theta^+$. In order to investigate the $3/2^+$ case, better experimental 
data of the resonances is needed.

The assignment of $J^P=3/2^-$ for exotic baryons seems reasonable also in a 
quark model especially when the narrow width of the $\Theta^+$ is to be 
explained\cite{Hosaka:2004bn}. The $(0s)^5$ configuration for the $3/2^-$ 
$\Theta^+$ is dominated by the $K^*N$ configuration\cite{Takeuchi:2004fv}, 
which however cannot be the decay channel, since the total masses of $K^*$ 
and $N$ is higher than the mass of $\Theta^+$. Hence we expect naturally (in
addition to a naive suppression mechanism due to the $d$-wave $KN$ decay) a 
strong suppression of the decay of the $\Theta^+$. 

The $3/2^-$ resonances of nonexotic quantum numbers have been also studied 
in various models of hadrons. A conventional quark model description with a 
$1p$ excitation of a single quark orbit has been successful 
qualitatively\cite{Isgur:1978xj}. Such three-quark states can couple with 
meson-baryon states which could be a source for the five- (or more-) quark 
content of the resonance. In the chiral unitary approach, $3/2^-$ states are
generated by $s$-wave scattering states of an octet meson and a decuplet
baryon\cite{Kolomeitsev:2003kt,Sarkar:2004jh}. By construction, the 
resulting resonances are largely dominated by five-quark content. These two 
approaches generate octet baryons which will eventually mix with the 
antidecuplet partners to generate the physical baryons. In other words, 
careful investigation of the octet states before mixing will provide further
information of the inner structure of the resonances\cite{Hyodo:2006uw}.

In the present phenomenological study, we have found that $J^P=3/2^-$ seems 
to fit the observed data. As we have known, other identifications have been 
also discussed in the literature. It is therefore important to determine the
quantum numbers of $\Theta^+$ in experiments, not only for the exotic 
particles but also for the baryon spectroscopy of nonexotic particles. Study
of high spin states in phenomenological models and calculations based on QCD
are strongly encouraged.

% % % % % % % % % % % % % % % % % 
\vspace{0.5cm}

We then studied the two-meson couplings of $\Theta^+$ for $J^P=1/2^+$ and 
$3/2^-$. The effective interaction Lagrangians for the
two-meson couplings were given, and these coupling constants were determined
based on the $\bm{8}$-$\overline{\bm{10}}$ representation mixing scheme, by 
using information of the $N^*\to \pi\pi N$ decays. These values were further
constrained in order to provide appropriate size of the self-energy of the 
$\Theta^+$. Finally, we have applied the effective interaction Lagrangians 
to the meson induced reactions $\piK$ and $\Kpi$.

We have found that there was an interference effect between the two 
amplitudes of the scalar and vector types, which could help to explain the 
very small cross section for the $\piK$ reaction observed at 
KEK\cite{Miwa:2006if}. In this case, reflecting the symmetry under
exchange of two amplitudes, large cross sections for $\Kpi$ reaction would 
be obtained as a consequence of the interference. The interference occurs in
both $1/2^+$ and $3/2^-$ cases.

In Table~\ref{tbl:summary}, we have summarized the results obtained in the 
present analysis. For a given set of coupling constants, the upper bound of 
the cross sections of the $\Kpi$ reactions were estimated by maximizing the 
interference effect. We observed that large cross sections of the order 
of millibarns for $\Kpi$ was obtained for the $1/2^+$ case, whereas the 
upper limit of the cross section was not very large for $3/2^-$ case. 
Therefore, if large cross sections are observed in the $\Kpi$ reaction, it 
would indicate $J^P=1/2^+$ for the $\Theta^+$.

For completeness, we would like to mention the case where the cross sections
for both $\piK$ and $\Kpi$ reactions are small. If the cross section of 
$\Kpi$ reaction is also small, it is not due to an interference effect, 
since the interference effect results in relatively large cross sections 
both for the two reactions. It could be explained by small coupling 
constants. For the $J^P=1/2^+$ case, both coupling constants can be zero 
within the experimental uncertainties. However, for the $3/2^-$ case, there 
is a lower limit for the $g_{3/2^-}^v$, which means that the lower limit is 
also imposed for the cross sections. We search for the set of coupling 
constants that provide the minimum value for the $\Kpi$ cross section, 
keeping a $\piK$ cross section to be less than $4.1 \mu$b. We obtain 
$\sigma_{\Kpi}\sim 58 \mu$b with $g_{3/2^-}^s=0.04$ and $g_{3/2^-}^v=0.18$. 
However, one should notice that the small coupling constants do not 
guarantee the dominance of two-meson coupling, and the Born terms and 
interference effect may play a role, which is beyond our present scope.

The present analysis provides an extension of effective interactions
obtained in Ref.~12 with representation mixing and the $J^P=3/2^-$ case. It 
is also interesting to apply the present extension to the study of the 
medium effect of $\Theta^+$\cite{Cabrera:2004yg} and the production of 
$\Theta^+$ hypernuclei\cite{Nagahiro:2004wu}. From the experimental point of
view, the cross section of $\Kpi$ reaction is of particular importance to 
the present results. To perform a better analysis for the two-meson 
coupling, more experimental data for three-body decays of nucleon resonances
are strongly desired.
 
%--------------------------------------------------------
\subsection*{Acknowledgements}
%--------------------------------------------------------

The authors would like to thank Koji Miwa, Takashi Nakano, Seung-il Nam, 
Eulogio Oset, Hiroshi Toki, Manuel J. Vicente Vacas, and Shi-Lin Zhu for 
useful discussions and comments.
One of the authors (T.H.) thanks to the Japan Society for the Promotion
of Science (JSPS) for support. This work supported in part by the Grant for 
Scientific Research [(C) No.17959600, T.H.] and [(C) No.16540252, A.H.] from
the Ministry of Education, Culture, Science and Technology, Japan.

\end{document}